 \documentclass[letterpaper, 10pt, conference, onecolumn]{ieeeconf}

\IEEEoverridecommandlockouts                              
\usepackage{epsfig}
\usepackage{url}
\usepackage{multicol}
\usepackage{amssymb}
\usepackage[tbtags]{amsmath}
\usepackage{latexsym}
\usepackage{euscript}
\usepackage{graphics}

\usepackage{fullpage}

\usepackage[tbtags]{amsmath} 
\usepackage{amssymb}  
\usepackage{verbatim} 

\DeclareMathOperator*{\argmax}{arg\,max}

\newcounter{actr}
{\begin{list}{(\alph{actr})}{\usecounter{actr}}}{\end{list}}

\newcounter{ictr}
{\begin{list}{(\roman{ictr})}{\usecounter{ictr}}}{\end{list}}

\newtheorem{theorem}{Theorem}
 \newtheorem{lemma}{Lemma}
 \newtheorem{definition}{Definition}
 \newtheorem{corollary}{Corollary}
 \newtheorem{proposition}{Proposition}

\newcommand{\qed}{\rule[0.1ex]{1.4ex}{1.6ex}}

\newcounter{psctr}
\newcounter{probctr}[psctr]

\DeclareMathAlphabet{\mathbsf}{OT1}{cmss}{bx}{n}
\DeclareMathAlphabet{\mathssf}{OT1}{cmss}{m}{sl}

\DeclareSymbolFont{bsfletters}{OT1}{cmss}{bx}{n}
\DeclareSymbolFont{ssfletters}{OT1}{cmss}{m}{n}
\DeclareMathSymbol{\bsfGamma}{0}{bsfletters}{'000}
\DeclareMathSymbol{\ssfGamma}{0}{ssfletters}{'000}
\DeclareMathSymbol{\bsfDelta}{0}{bsfletters}{'001}
\DeclareMathSymbol{\ssfDelta}{0}{ssfletters}{'001}
\DeclareMathSymbol{\bsfTheta}{0}{bsfletters}{'002}
\DeclareMathSymbol{\ssfTheta}{0}{ssfletters}{'002}
\DeclareMathSymbol{\bsfLambda}{0}{bsfletters}{'003}
\DeclareMathSymbol{\ssfLambda}{0}{ssfletters}{'003}
\DeclareMathSymbol{\bsfXi}{0}{bsfletters}{'004}
\DeclareMathSymbol{\ssfXi}{0}{ssfletters}{'004}
\DeclareMathSymbol{\bsfPi}{0}{bsfletters}{'005}
\DeclareMathSymbol{\ssfPi}{0}{ssfletters}{'005}
\DeclareMathSymbol{\bsfSigma}{0}{bsfletters}{'006}
\DeclareMathSymbol{\ssfSigma}{0}{ssfletters}{'006}
\DeclareMathSymbol{\bsfUpsilon}{0}{bsfletters}{'007}
\DeclareMathSymbol{\ssfUpsilon}{0}{ssfletters}{'007}
\DeclareMathSymbol{\bsfPhi}{0}{bsfletters}{'010}
\DeclareMathSymbol{\ssfPhi}{0}{ssfletters}{'010}
\DeclareMathSymbol{\bsfPsi}{0}{bsfletters}{'011}
\DeclareMathSymbol{\ssfPsi}{0}{ssfletters}{'011}
\DeclareMathSymbol{\bsfOmega}{0}{bsfletters}{'012}
\DeclareMathSymbol{\ssfOmega}{0}{ssfletters}{'012}

\newcommand{\rvC}{{\mathssf{C}}}    

\newcommand{\rva}{{\mathssf{a}}}    

\newcommand{\sva}{a}

\newcommand{\rvb}{{\mathssf{b}}}    
\newcommand{\rvc}{{\mathssf{c}}}    

\newcommand{\svc}{c}

\newcommand{\rvm}{{\mathssf{m}}}    

\newcommand{\rvs}{{\mathssf{s}}}    

\newcommand{\svs}{s}

\newcommand{\rvu}{{\mathssf{u}}}    
\newcommand{\svu}{u}

\newcommand{\rvw}{{\mathssf{w}}}    
\newcommand{\svw}{w}

\newcommand{\rvx}{{\mathssf{x}}}    

\newcommand{\svx}{x}            

\newcommand{\rvy}{{\mathssf{y}}}    

\newcommand{\svy}{y}

\newcommand{\rvz}{{\mathssf{z}}}    

\newcommand{\svz}{z}

\newcommand{\nchoosek}[2]{\left(\begin{array}{c}#1\\#2\end{array}\right)}
\begin{document}

\title{\LARGE \bf  On the rate distortion function of Bernoulli Gaussian sequences }
\author{Cheng Chang
\thanks{Cheng Chang is with Hewlett-Packard Labs, Palo Alto.
        {\tt\small Email:  cchang@eecs.berkeley.edu}}%
}

\maketitle

\begin{abstract}In this paper, we study the rate distortion
function of the i.i.d sequence of multiplications of a Bernoulli $p$
random variable and a gaussian random variable $\sim N(0,1)$. We use
a new technique in the derivation of the lower bound in which we
establish the duality between channel coding and lossy source coding
in the strong sense.  We improve the lower bound  on the rate
distortion function over the best known lower bound by
$p\log_2\frac{1}{p}$ if distortion $D$ is small. This has some
interesting implications on sparse signals where $p$ is small since
the known gap between the lower and upper bound is $H(p)$. This
improvement in the lower bound shows that the lower and upper bounds
are almost identical for sparse signals with small distortion
because $\lim\limits_{p\rightarrow
0}\frac{p\log_2\frac{1}{p}}{H(p)}=1$.
\end{abstract}

 \section{Bernoulli-Gaussian model and some obvious bounds on its rate distortion functions}
 \textbf{Notations:} in this paper we use $\rvx$, $\rvy$, $\rvu$ for random
variables and $\svx$, $\svy $, $\svu$ for the realization of the
random variables or constants. We denote by $\Pr\limits^\rvx( A)$
the probability of event $A$ under measure $\rvx$. We use bit and $\log_2$ in this paper.\\

 Consider a sequence of signals
$\rvx_1,\rvx_2,....\rvx_n$, where $\rvx_i$'s are zero most of the
time. When $\rvx_i$ is non-zero, it is an arbitrary real number. In
the signal processing literature, the signals $\rvx^n$ is called
sparse if most of them are zero. In their seminal work on
compressive sensing~\cite{Candes_Tao} and~\cite{Donoho}, Cand\`{e}s,
Tao and Donoho show that, to exactly reconstruct the sparse signals
$\rvx^n$, only a fraction of $n$ measurements are needed.
Furthermore, the reconstruction can be done by a linear programming
based efficient algorithm. In the compressed sensing literature, the
non-zero part of the sparse signals are arbitrary real numbers
without any statistical distribution assigned to them. Furthermore
the compressed sensing system tries to recover the signals $\rvx^n$
losslessly without distortion of the reconstructed signals. These
assumptions are not completely valid if the source statistics are
known to the coding system, more importantly, if the goal of the
sensing system is only to recover the data within a certain
distortion.  In the recent work by Fletcher etc.
~\cite{Fletcher_EURASIP,Fletcher_ICASSP,Fletcher_IEEE}, the .  What
is lacking in the previous study of this problem is a systematic
study of the information theoretic bounds on the rate distortion
functions of the sources. In this paper, we give both lower and
upper bounds on the rate distortion functions.

\subsection{Bernoulli-Gaussian random variable
$\Xi(p,\sigma^2)$}\label{sec.trivialbounds}

The information theoretic model of the ``sparse gaussian'' signals
is captured in the following what we call a Bernoulli-Gaussian
random variable.\vspace{0.1in}

\begin{definition}\label{def.bernoulli-gaussian} A random variable $\rvx$ is Bernoulli-Gaussian, denoted by
$\Xi(p,\sigma^2)$, if $\rvx=\rvb\times \rvs $, where $\rvs$ is a
Gaussian random variable with  mean  $0$ and variance $\sigma^2$,
$\rvs\sim N(0,\sigma^2)$, and $\rvb$ is a Bernoulli $p$ random
variable, $\Pr(\rvb=0)=1-p$ and $\Pr(\rvb=1)=p$, $p\in [0,1]$.
\end{definition}
\vspace{0.1in}

This random variable is a mixture of a continuous random variable
and a discrete random variable. This adds to the difficulties to
study the rate distortion functions of this random variable. The
main result of this paper is a lower bound and an upper bound on the
rate distortion functions of a sequence of independent random
variables with distribution $\Xi(p,\sigma^2)$. It will be clear soon
in Proposition~\ref{prop.equavalence} that we only need to study the
rate distortion functions for $\rvs\sim N(0,1)$, i.e. the rate
distortion functions for $\Xi(p,1)$. First, we review the definition
of rate distortion functions in both the average distortion  and
strong distortion sense.

\subsection{Review of the rate distortion theory}\label{sec.intro_R(D)}

  In the standard setup of rate distortion theory, the encoder maps
  $n$ i.i.d. random variables  $\rvx^n \in \mathcal X^n$, $\rvx\sim p_\rvx$, into $nR$ bits and then the
decoder reconstruct the original   signal within a certain
distortion. The encoder and decoder are denoted by $f_n$ and $g_n$
respectively:
$$f_n: \mathcal X^n \rightarrow \{0,1\}^{nR}\ \ and \
g_n: \{0,1\}^{nR} \rightarrow \hat {\mathcal X}^n,$$ and the
distortion is defined as $d(x^n, \hat
x^n)=\frac{1}{n}\sum\limits_{i=1}\limits^n d(x_i,\hat x_i)$.

\vspace{0.1in}
\begin{definition}{Rate distortion function (\cite{Cover}, pg.
341)}: the rate distortion function $R(D)$ is the infinimum of rates
$R$, such that $(R,D)$ is in the rate distortion region of the
source for a give distortion $D$. Where the rate distortion region
is the closure of achievable rate distortion pairs $(R,D)$ defined
as follows. $(R,D)$ is said to be achievable in the expected
distortion sense if there exists a sequence of $(2^{nR}, n)$ rate
codes $(f_n, g_n)$, such that
\begin{eqnarray}
\lim_{n\rightarrow \infty} E\left(d(\rvx^n,
g_n(f_n(\rvx^n)))\right)\leq D\label{eqn.defn_rd_average}
\end{eqnarray}

The strong sense of rate distortion function is defined similarly
with the following criteria for the codes: for all $\delta>0$
\begin{eqnarray}
\lim_{n\rightarrow \infty} \Pr\left(d(\rvx^n, g_n(f_n(\rvx^n))\geq
D+\delta \right)=0\label{eqn.defn_rd_strong}
\end{eqnarray}
where, in this paper, the distortion function $d(x^n, \hat
x^n)=\frac{1}{n}\sum\limits_{i=1}\limits^n (x_i-\hat x_i)^2$.
\end{definition}

It turns out that the rate distortions function for both the average
distortion and the strong distortion are the same for discrete
random variables Chapter 13.6~\cite{Cover}. We can generalize this
result easily to  continuous random variables whose variance is
finite and the probability density function satisfies the usual
regularity conditions. The proof can be carried out by quantizing
the probability density function and then by using the proof for
discrete random variables in~\cite{Cover}. A somewhat detailed
sketch of how this works is in Appendix~\ref{sec.appendix.strong}.

A good lossy coding system in the strong sense is not
\textit{necessarily} good in the expected distortion sense.
Considering the following example, a good lossy coder can miss the
distortion constraint for a subset $\Upsilon_n\subseteq\mathcal R^n$
with asymptotically $0$ measure, $\lim\limits_{n\rightarrow \infty}
\Pr\limits^\rvx(\Upsilon_n)=0$. However the good lossy coder can
\textit{intentionally} make the distortion on $\Upsilon_n$ no
smaller than  $\frac{2D}{\Pr\limits^\rvx(\Upsilon_n)}$, hence the
expected distortion is at least $2D$.

However it is easy to see that given a good lossy coding system in
the strong sense, we can easily make it also good in the expected
sense if the mean and variance of $\rvx$ are finite. We sketch the
proof in Appendix~\ref{sec.appendix.constructing}. So from now on,
when we say a lossy coding system is good in the strong sense, that
implies that the system is also good in the expected distortion
sense.

 \vspace{0.1in}
 The following lemma
characterizes the rate distortion function $R(D)$. \vspace{0.1in}
\begin{lemma}{Rate distortion theorem~\cite{Shannon}}:
\begin{eqnarray}
R(D)=\min_{p_{\hat \rvx| \rvx}: \sum\limits_{x,\hat x}p_\rvx(x)
p_{\hat \rvx| \rvx}(\hat x|x)d(x,\hat x)\leq D} I(\rvx; \hat
\rvx)\label{eqn.r(d)}.
\end{eqnarray}
\end{lemma}
\vspace{0.1in}

 \vspace{0.1in}
\begin{corollary}\label{cor.rate_gaussian}{Rate distortion theorem for Gaussian random
variables~\cite{Berger}:} for random variable $\rvx\sim N(0,
\sigma^2)$, the rate distortion function is:
\begin{eqnarray}
R(D,N(0, \sigma^2))= \{ \begin{array}{ccc}
\frac{1}{2}\log_2\frac{\sigma^2}{D}, & 0\leq D\leq \sigma^2,\\
0,& \ \ D> \sigma^2.
\end{array}
\end{eqnarray}
\end{corollary}
\vspace{0.1in}

 It is also shown that with the same variance and squared
distortion measure, Gaussian random variables requires the most bits
to be described. Both lower and upper bounds are given in Exercise 8
on Pg. 370~\cite{Cover}. The proof can be found in~\cite{Berger}.

 \vspace{0.1in}
\begin{corollary}\label{cor.rate_cover_bounds}{Rate distortion  bounds for continuous random variables
under square distortion measure (Exercise 8 on pg.
372~\cite{Cover}):} the rate distortion function $R(D)$ can be
bounded as:
\begin{eqnarray}
 h(\rvx) -\frac{D}{2}\log (2\pi e)\leq R(D) \leq \max\{\frac{1}{2}\log
 \frac{\sigma^2}{D},0\}\label{eqn.rate_cover_bounds}
 \end{eqnarray}
\end{corollary}

 \vspace{0.1in}
The lower bound in Corollary~\ref{cor.rate_cover_bounds} is known as
the Shannon lower bound in the literature~\cite{Cover}.

\subsection{Rate distortion function for $\Xi(p, \sigma^2)$}
The main goal of this paper is to derive an upper and a lower bound
on the rate distortion function $R(D)$ of the Bernoulli-Gaussian
random variable $\Xi(p,\sigma^2)$. We denote this quantity by $R(D,
\Xi(p,\sigma^2))$. We summarize some obvious properties of $R(D,
\Xi(p,\sigma^2))$ in the following four propositions. The proof is
in Appendix~\ref{sec.proof_upper}.

First we explain why we only need to study $R(D, \Xi(p, 1))$. We
write $R(D, \Xi(p,1))$ as $R(D,  p)$ in the rest of the paper and
investigate $R(D,p)$.

\vspace{0.1in}
\begin{proposition} \label{prop.equavalence}$R(D, \Xi(p,\sigma^2))=R(\frac{D}{\sigma^2}, \Xi(p, 1))$
\end{proposition}\vspace{0.1in}

\vspace{0.1in}
 From this point
on, we only investigate $ R({D}, \Xi(p, 1))$, simply written as
$R(D, p)$. Now we give three obvious bounds on the rate distortion
function $R(D, p)$.   \vspace{0.1in}
\begin{proposition}{Upper bound 1 on $R(D,p)$:}\label{prop.upper1}
\begin{eqnarray}
R(D,p)\leq H(p) +  p R(\frac{D}{p}, N(0,1))=  H(p) +  p R({D},
N(0,p))\label{eqn.upper1}
\end{eqnarray}
where $R(D, N(0,1))$ is the Gaussian rate distortion function for
$N(0,1)$, defined in Corollary~\ref{cor.rate_gaussian}.
\end{proposition}
 \vspace{0.1in}

 \begin{proposition}{Upper bound 2 on $R(D,p)$:}\label{prop.upper2}
\begin{eqnarray}
R(D,p)\leq    R(D, N(0,p))\label{eqn.upper2}
\end{eqnarray}
\end{proposition}
 \vspace{0.1in}

 \begin{proposition}{A lower bound  on $R(D,p)$:}\label{prop.lower_trivial}
\begin{eqnarray}
R(D,p)\geq   p R(\frac{D}{p}, N(0,1))= p R(D,
N(0,p))\label{eqn.lower_trivial}
\end{eqnarray}
\end{proposition}
 \vspace{0.1in}

We give a conceptually clear explanation of these three bounds. In
Proposition~\ref{prop.upper1}, we construct a very simple coding
system that first losslessly describe the  locations of the non-zero
elements of $\rvx^n\sim \Xi (p, 1)$, then lossily describe the value
of these non-zero elements using a Gaussian lossy coder. In
Proposition~\ref{prop.upper2}, we prove it by using the well known
fact that for continuous random variables, with the same variance
and distortion measure, Gaussian sequences require the highest rate.
The difficulty is that $\Xi(p,1)$ is not a continuous random
variable. We approximate it by a sequence of continuous random
variables whose rate distortion functions converge to that of
$\Xi(p, 1)$. In the proof of~\ref{prop.lower_trivial}, we reduce
  a Bernoulli-Gaussian sequence  to a Gaussian
sequence by letting the decoder know the non-zero locations for free
and derive \textit{a} lower bound of $R(D,p)$ from the Gaussian rate
distortion function.

The more rigorous proofs of these  bounds are in
Appendix~\ref{sec.proof_upper}. It is non trivial to bound the rate
distortion function of one random variable $\rvx$ by the rate
distortion function of another random variable $\rvy$.  To show that
$R(D, \rvx)\leq R(D,\rvy)$, the technique we use in the proofs for
the above four propositions is to \textit{construct} a good lossy
coding system for $\rvx$ from a good lossy coding system for $\rvy$
under the same rate-distortion constraint $R$ and $D$.

Among the three bounds described in
Proposition~\ref{prop.upper1},~\ref{prop.upper2}
and~\ref{prop.lower_trivial}, we find the lower bound the most
unsatisfactory. Shannon lower bound~\cite{Cover} does not apply to
the Bernoulli-Gaussian random $\Xi(p,1)$ variables because the
differential entropy of $\Xi(p,1)$ is negative infinity. This paper
is focused on deriving a more information-theoretically interesting
lower bound on $R(D,p)$. In the next several sections, we
investigate the lower bound problem. As a simple corollary of this
new lower bound, we give a close form lower bound on the rate
distortion function in~\ref{sec.discussions_num} that improves the
previous known result by $ p \log_2\frac{1}{p} $ in the high
resolution regime ($\frac{D}{p}\ll 1$).

\section{An improved lower bound on $R(D,p)$}
First, we reiterate the definition of a strong lossy source coding
system for a Bernoulli-Gaussian sequence $\rvx^n\sim \Xi(0,1)$ where
$\rvx=\rvb\times \rvs$ and $\rvb$ is a Bernoulli-$p$ random variable
while $\rvs\sim N(0,1)$ is a Gaussian random variable. A $(R,D)$
encoder-decoder sequence $f_n, g_n$ does the following,
\begin{eqnarray}
f_n: \mathcal R^n \rightarrow \{0,1\}^{nR}, \ \ f_n(x^n)=a^{nR} \ \
\mbox{ and } \ g_n: \{0,1\}^{nR} \rightarrow  {\mathcal R}^n,\ \
g_n(a^{nR})=\hat{\svx}^n\nonumber
\end{eqnarray}
from the definition of the rate distortion function in strong sense
defined in~(\ref{eqn.defn_rd_strong}), we have for all $\delta_1>0$:
\begin{eqnarray}
\Pr^\rvx\left(d(\rvx^n, \hat \rvx^n)\geq D+\delta_1\right)=
\Pr^\rvx\left(d(\rvx^n, g_n(f_n(\rvx^n)))\geq
D+\delta_1\right)=e_n(\delta_1) \mbox{  and  } \lim_{n\rightarrow
\infty} e_n(\delta_1)=0\label{eqn.codingpair_BG}.
\end{eqnarray}
\textbf{Recall that we can have a good lossy coder in both the
strong sense and the expected distortion sense according to the
discussions in Appendix~\ref{sec.appendix.constructing}. So we
assume the good coding system here $f_n, g_n$ is good in both
senses. }
\begin{eqnarray}
\mbox{So let  }E_\rvx\left(d(\rvx^n, \hat \rvx^n)\right)=
E_\rvx\left(d(\rvx^n, g_n(f_n(\rvx^n)))\right)=D+ \varsigma_n\mbox{,
then  } \lim_{n\rightarrow \infty}
\varsigma_n=0\label{eqn.codingpair_BG1}.
\end{eqnarray}

Notice that $\rvx^n=\rvb^n\times \rvs^n$, where the multiplication
$\times$ here is done entry by entry, so that if $\rvb_i=0$, the
value of $\rvs_i$ does not have any impact on $\rvx^n$. The output
of the encoder $f_n$ is a random variable that is a function of the
sequence $\rvx^n$, we write the output as $\rva^{nR}=f_n(\rvx^n)$.
our investigation of the rate distortion function relies on the
properties of the encoder output $\rva^{nR}$.

\vspace{0.1in}
\begin{figure}[htp]
\setlength{\unitlength}{3947sp}%
\begingroup\makeatletter\ifx\SetFigFont\undefined%
\gdef\SetFigFont#1#2#3#4#5{%
  \reset@font\fontsize{#1}{#2pt}%
  \fontfamily{#3}\fontseries{#4}\fontshape{#5}%
  \selectfont}%
\fi\endgroup%
\begin{picture}(6948,840)(2776,914)
{ \thinlines \put(4051,1139){\circle{212}}
}%
{ \put(3151,1139){\vector( 1, 0){750}}
}%
{ \put(4801,989){\framebox(1125,300){}}
}%
{ \put(4051,1514){\vector( 0,-1){225}}
}%
{ \put(4201,1139){\vector( 1, 0){525}}
}%
{ \put(7351,989){\framebox(1125,300){}}
}%
{ \put(8476,1139){\vector( 1, 0){450}}
}%
{ \put(6001,1139){\vector( 1, 0){1350}}
}%
\put(4951,1064){\makebox(0,0)[lb]{\smash{{\SetFigFont{12}{14.4}{\rmdefault}{\mddefault}{\updefault}{ Encoder $f_n$}%
}}}}
\put(3901,1664){\makebox(0,0)[lb]{\smash{{\SetFigFont{12}{14.4}{\rmdefault}{\mddefault}{\updefault}{ $\rvs^n$}%
}}}}
\put(7426,1064){\makebox(0,0)[lb]{\smash{{\SetFigFont{12}{14.4}{\rmdefault}{\mddefault}{\updefault}{ Decoder $g_n$}%
}}}}
\put(9001,1064){\makebox(0,0)[lb]{\smash{{\SetFigFont{12}{14.4}{\rmdefault}{\mddefault}{\updefault}{ $\hat \rvx^n$}%
}}}}
\put(4276,914){\makebox(0,0)[lb]{\smash{{\SetFigFont{12}{14.4}{\rmdefault}{\mddefault}{\updefault}{ $\rvx^n$}%
}}}}
\put(3906,1094){\makebox(0,0)[lb]{\smash{{\SetFigFont{12}{14.4}{\rmdefault}{\mddefault}{\updefault}{ $\times$}%
}}}}
\put(2776,1064){\makebox(0,0)[lb]{\smash{{\SetFigFont{12}{14.4}{\rmdefault}{\mddefault}{\updefault}{ $\rvb^n$}%
}}}}
\put(6901,1364){\makebox(0,0)[lb]{\smash{{\SetFigFont{12}{14.4}{\rmdefault}{\mddefault}{\updefault}{ $\rva^{nR}$}%
}}}}
\end{picture}%

\caption[ ]{ A lossy source coding system for Bernoulli-Gaussian
sequence $\rvx^n=\rvb^n\times \rvs^n$}
    \label{fig.BG_Lossycoding}
\end{figure}
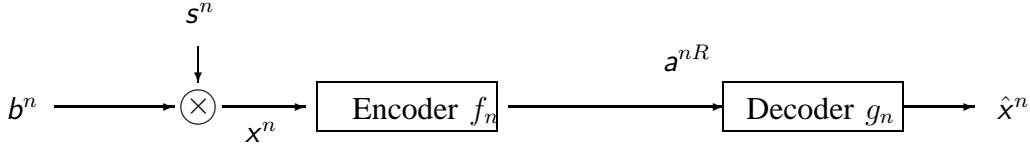
\vspace{0.1in}

 In Proposition~\ref{prop.lower_trivial}, the lower
bound is derived by letting a genie tell the decoder the non-zero
positions of the Bernoulli-Gaussian sequence, i.e. the $\rvb^n$ part
of $\rvx^n=\rvb^n\times \rvs^n$, and the rate is only counted for
the lossy source coding of the non-zero Gaussian subsequence $\tilde
\rvs^{1(\rvb^n)}$, where $1(\rvb^n)$ is the number of $1$'s in
sequence $\rvb^n$ and  $\tilde\rvs_i= \rvs_{l_i}$ if $\rvb_{l_i}=1$,
$i=1,2,...,1(\rvb^n)$. To tighten the lower bound in
Proposition~\ref{prop.lower_trivial}, we need to drop the genie who
let the decoder know the entirety of $\rvb^n$. In the following
several sections, we attempt to tighten up the lower bound by
investigating the information about $\rvb^n$ that \textit{has to} be
transmitted to the decoder.

First we summarize our main result in the following theorem.

\begin{theorem}{Main theorem: a new lower bound on the rate distortion
function $R(D,p)$ for Bernoulli-Gaussian random variable
$\Xi(p,1)$}\label{THM.mainresult} under distortion
 constraint $D$.
\begin{eqnarray}
&&R(D,p)\geq p R(D, N(0,p))+{\tilde R}\nonumber\\
\mbox{where } &&{\tilde R}= \max_{L \geq 0}\{\min_{U\geq L,
r\in[0,1-p]: T_1(L,U,r)\leq D } h(L,U,r) \}\label{eqn.maintheorem}
\end{eqnarray}
\begin{eqnarray}
\mbox{where   } h(L,U,r)=\{
\begin{array}{ccc}\nonumber
 (p\times\Pr(|\rvs|> U)+r)D(\frac{p\times\Pr(|\rvs|> U)}{p\times\Pr(|\rvs|> U)+r}\|p) &\mbox{,  if }\frac{p\times\Pr(|\rvs|> U)}{p\times\Pr(|\rvs|> U)+r} \geq p &\\
0   &\mbox{,  if } \frac{p\times\Pr(|\rvs|> U)}{p\times\Pr(|\rvs|>
U)+r}< p&
\end{array}
\end{eqnarray}
\end{theorem}
$\rvs\sim N(0,1)$ is a Gaussian random variable.\\

 \proof The theorem is a corollary of the
 Lemma~\ref{lemma.lowerboundtherate},~\ref{lemma:lowerboundI(a,s|b)},~\ref{lemma.lowerboundingI(a,b)}
 and~\ref{lemma.lossycoding_capacity}:
\begin{eqnarray}
R(D,p)&\geq&\frac{I(\rva^{nR}; \rvs^n|\rvb^n)+I(\rva^{nR}; \rvb^n)}{n}\label{eqn.mainttheorem.1}\\
&\geq& p R(D- (1-p)E[ \hat \rvx ^2|\rvb=0],
N(0,p)) +\frac{I(\rva^{nR}; \rvb^n)}{n}\label{eqn.mainttheorem.2}\\
&\geq& {p R(D- (1-p)E[ \hat \rvx ^2|\rvb=0],
N(0,p))} +   {\tilde R}  \label{eqn.mainttheorem.3}\\
&\geq& {p R(D, N(0,p))} +  {\tilde R} \label{eqn.mainttheorem.4}
\end{eqnarray}

(\ref{eqn.mainttheorem.1}) is proved in
Lemma~\ref{lemma.lowerboundtherate}. (\ref{eqn.mainttheorem.2}) is
proved in Lemma~\ref{lemma:lowerboundI(a,s|b)}.
(\ref{eqn.mainttheorem.3}) is proved in
Lemma~\ref{lemma.lowerboundingI(a,b)}
and~\ref{lemma.lossycoding_capacity}, $\tilde R$ is defined
in~(\ref{eqn.maintheorem}). (\ref{eqn.mainttheorem.4}) follows that
rate distortion function for Gaussian random variables $R(D,
N(0,p))$ is monotonically decreasing with $D$. \hfill $\blacksquare$
\vspace{0.1in}

There are four parts in our investigation. First in
Section~\ref{sec.sumofmutualinfo}, we lower bound the number of bits
$nR$ by the sum of two mutual information terms. The  first one is
the conditional mutual information between the output of the encoder
$\rva^{nR}$ and the Gaussian sequence $\rvs^n$ given the Bernoulli
sequence $\rvb^n$: $I(\rva^{nR}; \rvs^n|\rvb^n)$.   The second is
the mutual information between the output of the encoder $\rva^{nR}$
and the Bernoulli sequence $\rvb^n$: $I(\rva^{nR}; \rvb^n)$. Then in
Section~\ref{sec.boundingI(a,s)} we lower bound
 $I(\rva^{nR}; \rvs^n|\rvb^n)$ by using a simple argument similar to that
 in Proposition~\ref{prop.lower_trivial}. In
 Section~\ref{sec.boundingI(a,b)}, we lower bound $I(\rva^{nR};
 \rvb^n)$ by the capacity of the \textit{lossy coding channel},
 while the capacity of the channel is unspecified. In
 Section~\ref{sec.lowerboundingchannel}, we give \textit{a} lower
 bound of the channel capacity by using a random coding argument.
 Finally in Theorem~\ref{THM.mainresult}, we combine these bounds
 together to give a lower bound on the rate distortion function
 $R(D,p)$ for the Bernoulli-Gaussian random sequence $\Xi(p,1)$ under distortion
 constraint $D$. The investigation spans the next four sections in this paper.

\section{First step: lower bounding $nR$ by the sum of two mutual
information $I(\rva^{nR}; \rvb^n)+I(\rva^{nR};
\rvs^n|\rvb^n)$}\label{sec.sumofmutualinfo}

First we have the following simple lemma that tells us that the rate
is lower bounded by the sum of two mutual information terms
$I(\rva^{nR}; \rvb^n)+I(\rva^{nR}; \rvs^n)$ where $\rva^{nR}$ is the
output of the lossy encoder and $\rvb^n$ and $\rvs^n$ are the
Bernoulli sequence and the Gaussian sequence that generate the
Bernoulli-Gaussian $\rvx^n \sim  \Xi(p,1)$.\\

\begin{lemma}\label{lemma.lowerboundtherate} For a lossy coding system
 shown in Figure~\ref{fig.BG_Lossycoding}, the rate of the lossy
source coding system can be lower bounded as follows:
\begin{eqnarray}
nR \geq I(\rva^{nR}; \rvb^n)+I(\rva^{nR}; \rvs^n|\rvb^n)\nonumber
\end{eqnarray}
\end{lemma}
\vspace{0.1in}
 \proof The output of the encoder $\rva^{nR}\in
\{0,1\}^{nR}$, so the entropy of the random variable is upper
bounded by
\begin{eqnarray}
H(\rva^{nR} )\leq  nR \label{eqn.lowerboundRbyentropy}
\end{eqnarray}
Notice that $\rva^{nR}$ is a a function of $\rvx^n$, i.e. a function
of $\rvs^n$ and $\rvb^n$, so
\begin{eqnarray}
H(\rva^{nR} ) =  H(\rva^{nR} )- H(\rva^{nR}|\rvs^n, \rvb^n )
\label{eqn.lowerboundRbyentropy1}
\end{eqnarray}
Combining~(\ref{eqn.lowerboundRbyentropy})
and~(\ref{eqn.lowerboundRbyentropy1}), and notice that
$\rvb^n\bot\rvs^n$, we have:
\begin{eqnarray}
nR &\geq& H(\rva^{nR} )- H(\rva^{nR}|\rvs^n, \rvb^n )\nonumber\\
&=& I(\rva^{nR}; \rvs^n, \rvb^n)\nonumber\\
&= & I(\rva^{nR}; \rvb^n)+I(\rva^{nR}; \rvs^n|\rvb^n)
\label{eqn.lowerboundRbyentropy2}
\end{eqnarray}
where~(\ref{eqn.lowerboundRbyentropy2}) is true by the chain rule
for mutual information~\cite{Cover}. \hfill$\square$

\section{Lower bounding $I(\rva^{nR};
\rvs^n|\rvb^n)$, Proposition~\ref{prop.lower_trivial} revisited
}\label{sec.boundingI(a,s)} In this section we lower bound the
conditional mutual information term $I(\rva^{nR}; \rvs^n|\rvb^n)$ in
the lower bound of $nR$~(\ref{eqn.lowerboundRbyentropy2}). From
Proposition~\ref{prop.lower_trivial}, we know that letting a genie
tell the non-zero locations of $\rvx^n$ to the decoder,  the coding
system still needs at least $np R(D, N(0,p))$ bits to describe the
values of the non-zero entries of $\rvx^n$. In the proof of
Proposition~\ref{prop.lower_trivial}, like the proofs for other
propositions in Section~\ref{sec.trivialbounds}, we use the lossy
source coding system for the Bernoulli-Gaussian sequences to
construct a lossy source coding system  for a random sequence with
known rate distortion functions.

The proof here, however is trickier in the sense that we are not
bounding the rate distortion function $R(D,p)$, instead we only
bound the conditional mutual information $I(\rva^{nR};
\rvs^n|\rvb^n)$ which is  part of the rate. Hence we cannot
\textit{construct} a lossy coder for sequence with known rate
distortion using the lossy coder for the Bernoulli-Gaussian
sequence. Instead, we use the classical technique in~\cite{Cover}.

\vspace{0.1in}
\begin{lemma} {Lower bound on $I(\rva^{nR};
\rvs^n|\rvb^n)$}\label{lemma:lowerboundI(a,s|b)}
\begin{eqnarray}
I(\rva^{nR}; \rvs^n|\rvb^n)\geq np R(D- (1-p)E[ \hat \rvx
^2|\rvb=0], N(0,p))\label{eqn.lowerbound_conditional_MI}.
\end{eqnarray}
where
\begin{eqnarray}
E[ \hat \rvx ^2|\rvb=0]&=& \frac{1}{n} \sum_{i=1}^nE[ \hat \rvx_i
^2|\rvb_i=0]\nonumber\\
&=& \frac{1}{n}\left(\sum_{b^n }{\Pr(\rvb^n=b^n)}\sum_{i: b_i=0}^{}
   ( E[ \hat \rvx_{i}
^2|\rvb^n=b^n]) \right)
\end{eqnarray}
\end{lemma}

\vspace{0.1in}

\proof  The proof is similar to the lower bound proof for Gaussian
rate distortion function on Page 345~\cite{Cover}. First, notice
that the estimate $\hat \rvx^n= g_n(\rva^{nR})$ is  a function of
$\rva^n$. And the $\rva^{nR}=f_n(\rvx^n)=f_n(\rvb^n\times\rvs^n)$.
Hence we have the following Markov Chain:
\begin{eqnarray}
\rvb^n\times\rvs^n \rightarrow \rva^{nR}\rightarrow \hat \rvx^n
\label{eqn.markovchain}
\end{eqnarray}
From the data processing theorem~\cite{Cover}, we know that
$I(\rva^{nR}; \rvs^n|\rvb^n) \geq   I(\hat \rvx^n; \rvs^n|\rvb^n)$.
For a binary sequence $b^n\in \{0,1\}^n$, let
$1(b^n)=\sum\limits_{i=1}^n b_i$ be the number of $1$'s in $ b^n$.
$\rvb_i\in\{0,1\}$, so if $\rvb_i=0$ then $\hat \rvx^n$ and $\rvs_i$
are independent because in that case $\rvx_i=\rvb_i\times \rvs_i=0$
and $\rvs^n$ is i.i.d and $\hat\rvx^n$ is a deterministic function
of $\rvx^n$. Write $i_1,..., i_{1(b^n)}$ the non-zero positions of
$b^n$, and let $\mathcal I(b^n)=\{i_1,..., i_{1(b^n)}\}$, then
\begin{eqnarray}
I(\hat \rvx^n; \rvs^n|\rvb^n=b^n)=I(\hat \rvx^n;
\rvs_{i_1},...,\rvs_{i_{1(b^n)}}|\rvb^n=b^n)=I(\hat \rvx^n;
\rvs_{i_1,..., i_{1(b^n)}} )\label{eqn.condimi}.
\end{eqnarray}
Define the $\epsilon_1$-strong typical set $ B^n_{\epsilon_1}$ for
binary sequences:
\begin{eqnarray}
 B^n_{\epsilon_1}\triangleq\{b^n\in \{0,1\}^n:
|\frac{1(b^n)}{n}-p|\leq \epsilon_1\}.\nonumber
\end{eqnarray}
From the AEP~\cite{Cover}, let $\Pr(\rvb^n\notin
B^n_{\epsilon_1})=\upsilon_n$:
\begin{eqnarray}
\lim_{n\rightarrow \infty}
\upsilon_n=0\label{eqn.appendix.probabilityB1}
\end{eqnarray}

Now we have:
\begin{eqnarray}
I(\rva^{nR}; \rvs^n|\rvb^n)&\geq&  I(\hat \rvx^n;
\rvs^n|\rvb^n)\nonumber\\
&=&\sum_{b^n\in \{0,1\}^n}\Pr(\rvb^n=b^n)  I(\hat \rvx^n;
\rvs^n|\rvb^n=b^n)\label{eqn.condiMIdefinition}\\
&=&\sum_{b^n\in B^n_{\epsilon_1}}\Pr(\rvb^n=b^n)  I(\hat \rvx^n;
\rvs^n|\rvb^n=b^n)\label{eqn.condiMIAEP}\\
&=&\sum_{b^n\in B^n_{\epsilon_1}}\Pr(\rvb^n=b^n)  I(\hat \rvx^n;
\rvs_{i_1},...,\rvs_{i_{1(b^n)}}|\rvb^n=b^n)\label{eqn.usingcondimi}\\
&=& \sum_{b^n\in B^n_{\epsilon_1}}\Pr(\rvb^n=b^n)
\left(H(\rvs_{i_1,...,
i_{1(b^n)}}|\rvb^n=b^n)- H(\rvs_{i_1,..., i_{1(b^n)}}|\hat \rvx^n,\rvb^n=b^n)\right)\nonumber\\
&\geq & \sum_{b^n\in B^n_{\epsilon_1}}\Pr(\rvb^n=b^n)
\left(\sum_{j=1}^{{1(b^n)}}H(\rvs_{i_j})- \sum_{j=1}^{{1(b^n)}}H(\rvs_{i_j}|\hat \rvx^n,\rvb^n=b^n)\right)\label{eqn.lowerboundmutal.1}\\
&= & \sum_{b^n\in B^n_{\epsilon_1}}\Pr(\rvb^n=b^n)
\left(\sum_{j=1}^{{1(b^n)}}H(\rvs_{i_j})- \sum_{j=1}^{{1(b^n)}}H(\rvs_{i_j}-\hat \rvx_{i_j}|\hat \rvx^n,\rvb^n=b^n)\right)\nonumber\\
&\geq & \sum_{b^n\in B^n_{\epsilon_1}}\Pr(\rvb^n=b^n)
\left(\sum_{j=1}^{{1(b^n)}}H(\rvs_{i_j})-
\sum_{j=1}^{{1(b^n)}}H(\rvs_{i_j}-\hat \rvx_{i_j}
|\rvb^n=b^n)\right)\label{eqn.lowerbounding_I(a,s)}
\end{eqnarray}
(\ref{eqn.condiMIdefinition}) follows the definition of conditional
mutual information, (\ref{eqn.condiMIAEP}) is true because mutual
information is non-negative and~(\ref{eqn.usingcondimi})
follows~(\ref{eqn.condimi}). (\ref{eqn.lowerboundmutal.1}) is true
because $\rvs^n$ is i.i.d and independent of $\rvb^n$. The rest are
obvious. $\rvs_i\sim N(0,1)$, so $H(\rvs_{i})=\frac{1}{2}\log(2\pi
e)$. According Theorem 9.6.5 in~\cite{Cover}, Gaussian random
variables maximize the entropy over all distributions with thes ame
covariance, so:
$$H(\rvs_{i_j}-\hat \rvx_{i_j} |\rvb^n=b^n)\leq H(N(0,E[(\rvs_{i_j}-\hat \rvx_{i_j} )^2|\rvb^n=b^n])
=\frac{1}{2}\log(2\pi e E[(\rvs_{i_j}-\hat \rvx_{i_j}
)^2|\rvb^n=b^n]).$$
Now~(\ref{eqn.lowerbounding_I(a,s)}) becomes:

\begin{eqnarray}
I(\rva^{nR}; \rvs^n|\rvb^n)  &\geq & \sum_{b^n\in
B^n_{\epsilon_1}}\Pr(\rvb^n=b^n) \left(\sum_{j=1}^{{1(b^n)}}
\frac{1}{2}\log(2\pi e) - \sum_{j=1}^{{1(b^n)}} \frac{1}{2}\log(2\pi
e E[(\rvs_{i_j}-\hat \rvx_{i_j}
)^2|\rvb^n=b^n]) \right)\nonumber\\
&=&  \sum_{b^n\in B^n_{\epsilon_1}}\Pr(\rvb^n=b^n)
\left(-\sum_{j=1}^{{1(b^n)}}   \frac{1}{2}\log( E[(\rvs_{i_j}-\hat
\rvx_{i_j}
)^2|\rvb^n=b^n]) \right)\nonumber\\
&=&  -\frac{1}{2}\sum_{b^n\in
B^n_{\epsilon_1}}\sum_{j=1}^{{1(b^n)}}\Pr(\rvb^n=b^n) \left(
   \log( E[(\rvs_{i_j}-\hat \rvx_{i_j}
)^2|\rvb^n=b^n]) \right)\nonumber\\
&=&  -\frac{1}{2}\left(\sum_{b^n\in
B^n_{\epsilon_1}}\sum_{j=1}^{{1(b^n)}}\frac{\Pr(\rvb^n=b^n)}{\sum\limits_{b^n\in
B^n_{\epsilon_1}}\sum\limits_{j=1}^{{1(b^n)}}\Pr(\rvb^n=b^n)}
   \log( E[(\rvs_{i_j}-\hat \rvx_{i_j}
)^2|\rvb^n=b^n]) \right)\times \nonumber\\
&&\ \ \ \ \ \ \ \left( {\sum_{b^n\in
B^n_{\epsilon_1}}\sum_{j=1}^{{1(b^n)}}\Pr(\rvb^n=b^n)}\right)\nonumber\\
&\geq&  -\frac{1}{2}\log\left(\sum_{b^n\in
B^n_{\epsilon_1}}\sum_{j=1}^{{1(b^n)}}\frac{\Pr(\rvb^n=b^n)}{\sum\limits_{b^n\in
B^n_{\epsilon_1}}\sum\limits_{j=1}^{{1(b^n)}}\Pr(\rvb^n=b^n)}
   ( E[(\rvs_{i_j}-\hat \rvx_{i_j}
)^2|\rvb^n=b^n]) \right)\times \nonumber\\
&&\ \ \ \ \ \ \ \left( {\sum_{b^n\in
B^n_{\epsilon_1}}\sum_{j=1}^{{1(b^n)}}\Pr(\rvb^n=b^n)}\right)
  \label{eqn.convex_I(a,s)}
\end{eqnarray}
(\ref{eqn.convex_I(a,s)}) follows the fact that $-\log (\cdot)$ is
convex $\bigcup$. We bound the two terms as follows, first:
\begin{eqnarray}
\left( {\sum_{b^n\in
B^n_{\epsilon_1}}\sum_{j=1}^{{1(b^n)}}\Pr(\rvb^n=b^n)}\right)&=&
\left( {\sum_{b^n\in
B^n_{\epsilon_1}} 1(b^n) \Pr(\rvb^n=b^n)}\right) \nonumber\\
&\geq& \left( {\sum_{b^n\in
B^n_{\epsilon_1}} n(p-\epsilon_1)\Pr(\rvb^n=b^n)}\right) \nonumber\\
&\geq& n(p-\epsilon_1)(1-\upsilon_n)\label{eqn.bounding_I(a,s)_0.5}
\end{eqnarray}
Before bounding the other term, we have the following observation:

\begin{eqnarray}
&&\left(\sum_{b^n\in B^n_{\epsilon_1}}\sum_{j=1}^{{1(b^n)}}
{\Pr(\rvb^n=b^n)}
   ( E[(\rvs_{i_j}-\hat \rvx_{i_j}
)^2|\rvb^n=b^n]) \right)\nonumber\\
 &=& \left(\sum_{b^n\in
B^n_{\epsilon_1}}\sum_{j=1}^{{1(b^n)}} {\Pr(\rvb^n=b^n)}
   ( E[(\rvx_{i_j}-\hat \rvx_{i_j}
)^2|\rvb^n=b^n]) \right)\nonumber\\
 &\leq & \left(\sum_{b^n}\sum_{j=1}^{{1(b^n)}} {\Pr(\rvb^n=b^n)}
   ( E[(\rvx_{i_j}-\hat \rvx_{i_j}
)^2|\rvb^n=b^n]) \right)\nonumber\\
 &\leq& n(D+\varsigma_n)-\left(\sum_{b^n }\sum_{i\notin\mathcal I(b^n)}^{} {\Pr(\rvb^n=b^n)}
   ( E[(\rvx_{i}-\hat \rvx_{i}
)^2|\rvb^n=b^n]) \right)\label{eqn.bounding_I(a,s)_0.9}\\
 &\leq& n(D+\varsigma_n)-\left(\sum_{b^n }\sum_{i\notin\mathcal I(b^n)}^{} {\Pr(\rvb^n=b^n)}
   ( E[ \hat \rvx_{i}
^2|\rvb^n=b^n]) \right)\nonumber\\
&=& n(D+\varsigma_n)- n E[ \hat \rvx
^2|\rvb=0]\label{eqn.bounding_I(a,s)_1}
\end{eqnarray}
where $\mathcal I(b^n)=\{i_1,..., i_{1(b^n)}\}$  and
$\varsigma_n\rightarrow 0$ as $n$ goes to
infinity,~(\ref{eqn.bounding_I(a,s)_0.9}) follows the fact that
$f_n, g_n$ is good in the expected distortion sense as
well~(\ref{eqn.codingpair_BG1}). So the first term
in~(\ref{eqn.convex_I(a,s)}) can be lower bounded as follows,
combining~(\ref{eqn.bounding_I(a,s)_0.5})
and~(\ref{eqn.bounding_I(a,s)_1}):
\begin{eqnarray}
-\frac{1}{2}\log\left(\sum_{b^n\in
B^n_{\epsilon_1}}\sum_{j=1}^{{1(b^n)}}\frac{\Pr(\rvb^n=b^n)}{\sum\limits_{b^n\in
B^n_{\epsilon_1}}\sum\limits_{j=1}^{{1(b^n)}}\Pr(\rvb^n=b^n)}
   ( E[(\rvs_{i_j}-\hat \rvx_{i_j}
)^2|\rvb^n=b^n]) \right)\geq-\frac{1}{2}\log\left(\frac{(D- E[ \hat
\rvx
^2|\rvb=0]+\varsigma_n)}{(p-\epsilon_1)(1-\upsilon_n)}\right)\label{eqn.bounding_I(a,s)_2}
\end{eqnarray}

first notice that we are lower bounding a conditional mutual
information $I(\rva^{nR}; \rvs^n|\rvb^n) $ which is non-negative, so
we assume the first term being positive or else we lower bound the
conditional mutual information by $0$, so
substituting~(\ref{eqn.bounding_I(a,s)_0.5})
and~(\ref{eqn.bounding_I(a,s)_2}) into~(\ref{eqn.convex_I(a,s)}), we
have:

\begin{eqnarray}
I(\rva^{nR}; \rvs^n|\rvb^n)  &\geq &
n(p-\epsilon_1)(1-\upsilon_n)\max\{0,
\log\left(\frac{(p-\epsilon_1)(1-\upsilon_n)}{(D- (1-p)E[ \hat \rvx
^2|\rvb=0]+\varsigma_n)}\right)\}
\end{eqnarray}

Notice that $\epsilon_1$ is an arbitrary positive real number, and
both $\upsilon_n$ and $\varsigma_n$ goes to zero as $n$ goes to
infinity, so we just showed that

\begin{eqnarray}
I(\rva^{nR}; \rvs^n|\rvb^n)  &\geq & np\times\max\{0,
\log\left(\frac{p}{D- (1-p)E[ \hat \rvx ^2|\rvb=0]}\right)\}=np
R(D-(1-p) E[ \hat \rvx ^2|\rvb=0], N(0,p))\nonumber
\end{eqnarray}

The lemma is proved.  \hfill $\square$ \vspace{0.1in}

As a trivial corollary of Lemma~\ref{lemma.lowerboundtherate} and
Lemma~\ref{lemma:lowerboundI(a,s|b)}, we have:
\begin{eqnarray}
nR \geq I(\rva^{nR}; \rvb^n)+I(\rva^{nR}; \rvs^n|\rvb^n)\geq np R(D-
(1-p)E[ \hat \rvx ^2|\rvb=0], N(0,p))\geq np R(D, N(0,p))\nonumber
\end{eqnarray}
This also proves Proposition~\ref{prop.lower_trivial}.

\section{Lower bounding $I(\rva^{nR},
\rvb^n)$ by the randomized channel capacity of a lossy
compressor}\label{sec.boundingI(a,b)}

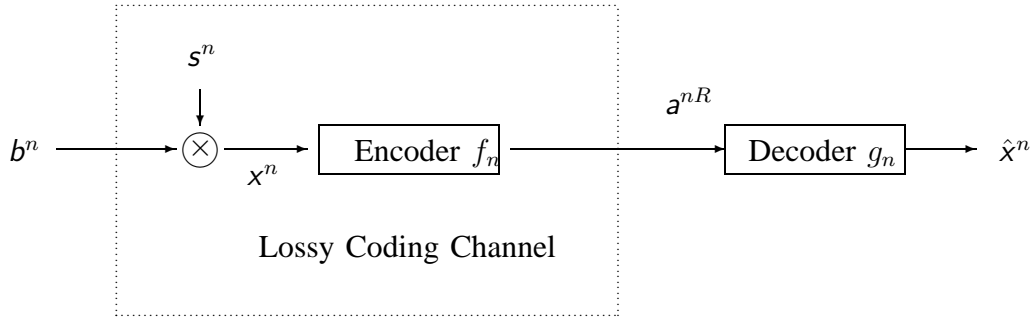
\begin{figure}[htp]
\setlength{\unitlength}{3947sp}%
\begingroup\makeatletter\ifx\SetFigFont\undefined%
\gdef\SetFigFont#1#2#3#4#5{%
  \reset@font\fontsize{#1}{#2pt}%
  \fontfamily{#3}\fontseries{#4}\fontshape{#5}%
  \selectfont}%
\fi\endgroup%
\begin{picture}(6948,1974)(2776,77)
{ \thinlines \put(4051,1139){\circle{212}}
}%
{ \put(3151,1139){\vector( 1, 0){750}}
}%
{ \put(4801,989){\framebox(1125,300){}}
}%
{ \put(4051,1514){\vector( 0,-1){225}}
}%
{ \put(4201,1139){\vector( 1, 0){525}}
}%
{ \put(7351,989){\framebox(1125,300){}}
}%
{ \put(8476,1139){\vector( 1, 0){450}}
}%
{ \put(6001,1139){\vector( 1, 0){1350}}
}%
{ \multiput(3526,
89)(36.00000,0.00000){88}{\makebox(1.6667,11.6667){\SetFigFont{5}{6}{\rmdefault}{\mddefault}{\updefault}.}}
\multiput(3526,2039)(36.00000,0.00000){88}{\makebox(1.6667,11.6667){\SetFigFont{5}{6}{\rmdefault}{\mddefault}{\updefault}.}}
\multiput(3526,
89)(0.00000,35){56}{\makebox(1.6667,11.6667){\SetFigFont{5}{6}{\rmdefault}{\mddefault}{\updefault}.}}
\multiput(6676,
89)(0.00000,35){56}{\makebox(1.6667,11.6667){\SetFigFont{5}{6}{\rmdefault}{\mddefault}{\updefault}.}}
}%
\put(4951,1064){\makebox(0,0)[lb]{\smash{{\SetFigFont{12}{14.4}{\rmdefault}{\mddefault}{\updefault}{ Encoder $f_n$}%
}}}}
\put(3901,1664){\makebox(0,0)[lb]{\smash{{\SetFigFont{12}{14.4}{\rmdefault}{\mddefault}{\updefault}{ $\rvs^n$}%
}}}}
\put(7426,1064){\makebox(0,0)[lb]{\smash{{\SetFigFont{12}{14.4}{\rmdefault}{\mddefault}{\updefault}{ Decoder $g_n$}%
}}}}
\put(9001,1064){\makebox(0,0)[lb]{\smash{{\SetFigFont{12}{14.4}{\rmdefault}{\mddefault}{\updefault}{ $\hat \rvx^n$}%
}}}}
\put(4276,914){\makebox(0,0)[lb]{\smash{{\SetFigFont{12}{14.4}{\rmdefault}{\mddefault}{\updefault}{ $\rvx^n$}%
}}}}
\put(3906,1100){\makebox(0,0)[lb]{\smash{{\SetFigFont{12}{14.4}{\rmdefault}{\mddefault}{\updefault}{ $\times$}%
}}}}
\put(2776,1064){\makebox(0,0)[lb]{\smash{{\SetFigFont{12}{14.4}{\rmdefault}{\mddefault}{\updefault}{ $\rvb^n$}%
}}}}
\put(6901,1364){\makebox(0,0)[lb]{\smash{{\SetFigFont{12}{14.4}{\rmdefault}{\mddefault}{\updefault}{ $\rva^{nR}$}%
}}}}
\put(4351,464){\makebox(0,0)[lb]{\smash{{\SetFigFont{12}{14.4}{\rmdefault}{\mddefault}{\updefault}{ Lossy Coding Channel }%
}}}}
\end{picture}%

\caption[ ]{ A ``lossy coding'' channel derived from the lossy
coding system for Bernoulli-Gaussian sequence $\rvx^n=\rvb^n\times
\rvs^n$, }
    \label{fig.BG_Lossycoding_channel}
\end{figure}

In this section we give a lower bound on the mutual information
$I(\rva^{nR}; \rvb^n)$ from a channel capacity perspective. This is
partly inspired by the seminal work in~\cite{Sahai_Agarwal}. First
we have another look at the whole lossy coding system in
Figure~\ref{fig.BG_Lossycoding}, we single out the binary randomness
$\rvb^n$ and make the rest of the system a ``lossy coding channel''
as shown in Figure~\ref{fig.BG_Lossycoding_channel}. The channel
input is a binary sequence $b^n\in \{0,1\}^n$, and the channel
output is $a^{nR}\in \{0,1\}^{nR}$. What the channel does is to
first multiply $b^n$ by a Gaussian random sequence $\rvs^n$ and then
send it to a good lossy encoder $f_n$. The output is the output of
the lossy coding encoder $f_n$.

Notice that this is not a standard communication channel. It is in
some sense a arbitrarily varying channel. The constraint on the
channel is such that the lossy coder pair $f_n, g_n$ is good in both
the strong and expected distortion sense. \textbf{The goal in this
section is to lower bound the mutual information $I(\rva^{nR},
\rvb^n)$  by the number bits (channel capacity) that can be reliably
communicated across the channel in average over a randomized
codebook.}

More interestingly, the input sequence $\rvb^n$ obeys the statistics
of a Bernoulli process with non-zero probability $p$. So it will be
soon obvious that we need to investigate the channel capacity for
the randomized codebooks where each code word is chosen according to
its probability under i.i.d Bernoulli-$p$.

\vspace{0.1in}
\vspace{0.1in}

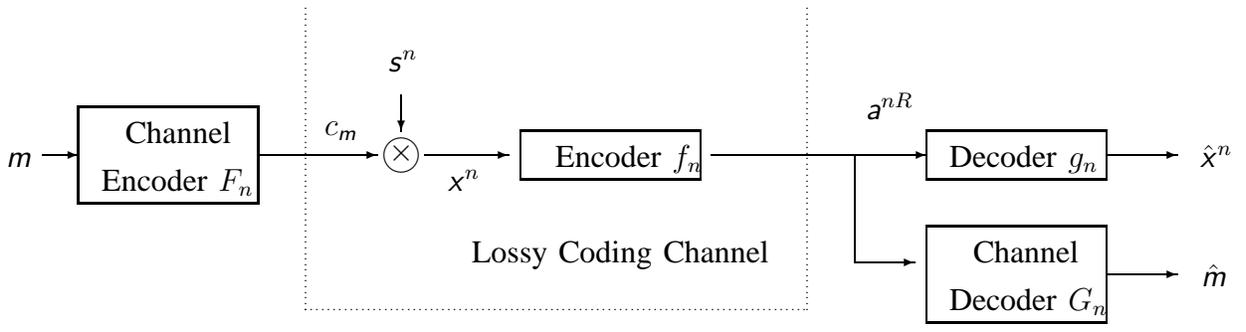
\begin{figure}[htp]
\setlength{\unitlength}{3947sp}%
\begingroup\makeatletter\ifx\SetFigFont\undefined%
\gdef\SetFigFont#1#2#3#4#5{%
  \reset@font\fontsize{#1}{#2pt}%
  \fontfamily{#3}\fontseries{#4}\fontshape{#5}%
  \selectfont}%
\fi\endgroup%
\begin{picture}(8298,2049)(1501,77)
{ \thinlines \put(4051,1139){\circle{212}}
}%
{ \put(3151,1139){\vector( 1, 0){750}}
}%
{ \put(4801,989){\framebox(1125,300){}}
}%
{ \put(4051,1514){\vector( 0,-1){225}}
}%
{ \put(4201,1139){\vector( 1, 0){525}}
}%
{ \put(7351,989){\framebox(1125,300){}}
}%
{ \put(8476,1139){\vector( 1, 0){450}}
}%
{ \put(6001,1139){\vector( 1, 0){1350}}
}%
{
\multiput(3451,164)(36.00000,0.00000){88}{\makebox(1.6667,11.6667){\SetFigFont{5}{6}{\rmdefault}{\mddefault}{\updefault}.}}
\multiput(3451,2114)(36.00000,0.00000){88}{\makebox(1.6667,11.6667){\SetFigFont{5}{6}{\rmdefault}{\mddefault}{\updefault}.}}
\multiput(3451,164)(0.00000,35){55}{\makebox(1.6667,11.6667){\SetFigFont{5}{6}{\rmdefault}{\mddefault}{\updefault}.}}
\multiput(6601,164)(0.00000,35){55}{\makebox(1.6667,11.6667){\SetFigFont{5}{6}{\rmdefault}{\mddefault}{\updefault}.}}
}%
{ \put(6901,1139){\line( 0,-1){675}} \put(6901,464){\vector( 1,
0){375}}
}%
{ \put(2026,839){\framebox(1125,600){}}
}%
{ \put(7351, 89){\framebox(1125,600){}}
}%
{ \put(8476,389){\vector( 1, 0){450}}
}%
{ \put(1801,1139){\vector( 1, 0){225}}
}%
\put(4951,1064){\makebox(0,0)[lb]{\smash{{\SetFigFont{12}{14.4}{\rmdefault}{\mddefault}{\updefault}{ Encoder $f_n$}%
}}}}
\put(3901,1664){\makebox(0,0)[lb]{\smash{{\SetFigFont{12}{14.4}{\rmdefault}{\mddefault}{\updefault}{ $\rvs^n$}%
}}}}
\put(7426,1064){\makebox(0,0)[lb]{\smash{{\SetFigFont{12}{14.4}{\rmdefault}{\mddefault}{\updefault}{ Decoder $g_n$}%
}}}}
\put(9001,1064){\makebox(0,0)[lb]{\smash{{\SetFigFont{12}{14.4}{\rmdefault}{\mddefault}{\updefault}{ $\hat \rvx^n$}%
}}}}
\put(4276,914){\makebox(0,0)[lb]{\smash{{\SetFigFont{12}{14.4}{\rmdefault}{\mddefault}{\updefault}{ $\rvx^n$}%
}}}}
\put(3906,1100){\makebox(0,0)[lb]{\smash{{\SetFigFont{12}{14.4}{\rmdefault}{\mddefault}{\updefault}{ $\times$}%
}}}}
\put(6901,1364){\makebox(0,0)[lb]{\smash{{\SetFigFont{12}{14.4}{\rmdefault}{\mddefault}{\updefault}{ $\rva^{nR}$}%
}}}}
\put(4426,464){\makebox(0,0)[lb]{\smash{{\SetFigFont{12}{14.4}{\rmdefault}{\mddefault}{\updefault}{ Lossy Coding Channel }%
}}}}
\put(7426,164){\makebox(0,0)[lb]{\smash{{\SetFigFont{12}{14.4}{\rmdefault}{\mddefault}{\updefault}{ Decoder $G_n$}%
}}}}
\put(2101,914){\makebox(0,0)[lb]{\smash{{\SetFigFont{12}{14.4}{\rmdefault}{\mddefault}{\updefault}{ Encoder $F_n$}%
}}}}
\put(2251,1214){\makebox(0,0)[lb]{\smash{{\SetFigFont{12}{14.4}{\rmdefault}{\mddefault}{\updefault}{ Channel}%
}}}}
\put(7576,464){\makebox(0,0)[lb]{\smash{{\SetFigFont{12}{14.4}{\rmdefault}{\mddefault}{\updefault}{ Channel}%
}}}}
\put(1501,1064){\makebox(0,0)[lb]{\smash{{\SetFigFont{12}{14.4}{\rmdefault}{\mddefault}{\updefault}{ $\rvm$}%
}}}}
\put(9076,314){\makebox(0,0)[lb]{\smash{{\SetFigFont{12}{14.4}{\rmdefault}{\mddefault}{\updefault}{$\hat \rvm$}%
}}}}
\put(3501,1264){\makebox(0,0)[lb]{\smash{{\SetFigFont{12}{14.4}{\rmdefault}{\mddefault}{\updefault}{ $c_\rvm$}%
}}}}

\end{picture}%

\caption[ ]{ A channel coding system for the ``lossy coding''
channel }
    \label{fig.BG_Lossycoding_channel_coding}
\end{figure}

As shown in Figure~\ref{fig.BG_Lossycoding_channel_coding}, we have
a channel coding problem.  A message $\rvm$ is a random variable
uniformly distributed on $\{1,2,...,2^{nR}\}$. The constraint on the
channel encoder $F_n$ is that the code word   $b^n$ is chosen for
message $m$ with probability $$p^{1(b^n)}(1-p)^{n-1(b^n)},$$
  where $1(b^n)$ is the number of
$1$'s in sequence $b^n$, this will be explained in details in
Definition~\ref{def.randomzied_capacity}. The constraint on the
\textit{lossy coding channel} is such that the estimate of the
Bernoulli Gaussian random sequence $\rvx^n=\rvb^n\times \rvs^n$,
through the lossy coding system $f_n,g_n$: $\hat \rvx^n$ is within a
distortion $D+\delta_1$ of the true sequence $\rvx^n$ with
probability $1$ for all $\delta_1>0$ asymptotically. \vspace{0.1in}
\vspace{0.1in} Before giving the lemma on the lower bound of the
mutual information $I(\rva^{nR}; \rvb^n)$, we give the following
definition  of randomized channel capacity for the lossy source
channel. \vspace{0.1in}

\begin{definition}\label{def.randomzied_capacity}{ Randomized channel capacity for the lossy source
channel is written as $\tilde R_p$}\footnote{Note: in this section
we use $\tilde R$ to denote the channel capacity of the lossy coding
channel. This is not the rate of the lossy coding system $R$.}: let
$\mathcal B_n=\{0,1\}^n$, let $\mathcal C(n)$ be the codebook set of
rate $\tilde R$: $ \mathcal C(n)= \mathcal B_n^{2^{n\tilde R}} $ is
the set product of $2^{n\tilde R}$ many $\mathcal B_n$'s: $ \mathcal
 B_n\times\mathcal  B_n\times...\times \mathcal  B_n$, a
codebook $  C\in \mathcal C(n)$, $C=(c_1,c_2,...c_{2^{n\tilde R}})$
is such that the codeword for message $m$, $m=1,2,...2^{n\tilde R}$,
is the $i$'th entry of $C$: $c_m$. From the definition
$c_m\in\mathcal B_n$ for all $n$.  We let $\rvC_p$ be a random
variable distributed on $\mathcal C(n)$, such that a codebook
$C=(c_1,c_2,...c_{2^{n\tilde R}})\in \mathcal C(n) $ is chosen as
the codebook, i.e. $\rvC_p=C$ with the following probability:
\begin{eqnarray}
\Pr(\rvC_p=C)=\prod_{m=1}^{2^{n\tilde R}}
p^{1(c_m)}(1-p)^{n-1(c_m)}\label{eqn.randomized_distribution}
\end{eqnarray}

the average error probability of the randomized coding with uniform
distributed $\rvC_p$ is defined as:
\begin{eqnarray}
e_{p,n}(\tilde R)&= &\sum_{C\in \mathcal B_n ^{2^{n\tilde
R}}}\Pr(\rvC_p=C)\left(\frac{1}{2^{n\tilde R}}\sum_{m=1}^{2^{n\tilde
R}}\Pr^{\rvs}(m\neq \hat m (\sva^{nR}(c_m\times
\rvs^n)))\right)\nonumber\\
&=& \sum_{C\in \mathcal B_n ^{2^{n\tilde
R}}}\Pr(\rvC_p=C)\left(\Pr(\rvm \neq \hat
\rvm(\rva^{nR})|\rvC_p=C)\right)\label{eqn.error_randomizedcode}
\end{eqnarray}
where the error probability is over all codebooks $ \mathcal
C(n)=\mathcal B_q ^{2^{n\tilde R}}$ with distribution defined
in~(\ref{eqn.randomized_distribution}) and all messages
$m\in\{1,2,...,2^{n\tilde R}\}$, i.e. the random variable $\rvm$ is
uniformly distributed in~(\ref{eqn.error_randomizedcode}). Notice
that in Figure~\ref{fig.BG_Lossycoding_channel_coding}, a codebook
$C$ is chosen and known to both the encoder and the decoder. The
output from the channel encoder is $F_n(m)=c_m$, the output from the
lossy encoder is a random sequence $f_n(c_m\times
\rvs^n)=\sva^{nR}(c_m\times \rvs^n)$, and the estimate of $m$ is
$\hat m(\sva^{nR}(c_m\times \rvs^n))=G_n(\sva^{nR}(c_m\times
\rvs^n)))$.

The randomized channel capacity  for the lossy coding system
$f_n,g_n$ is $\tilde R_{ p}$, if for all $\tilde R<\tilde R_{ p}$,
there exists a channel decoder $ G_n$, such that the average error
goes to zero as $n$ goes to infinity:
$$\lim\limits_{ n\rightarrow \infty}e_{p,n}(\tilde R)=0, \mbox{ equivalently: } \tilde R_p=\sup_{\lim\limits_{ n\rightarrow \infty}e_{p,n}(\tilde R)=0}\{\tilde R\}.$$

\end{definition}

\vspace{0.1in}
 The following lemma summarizes the main result in
this section.

\vspace{0.1in}
\begin{lemma}\label{lemma.lowerboundingI(a,b)}{Lower bounding the mutual information  $I(\rva^{nR},
\rvb^n) $ by the randomized capacity}: for any $\epsilon>0$ the
mutual information is lower bounded by the minimum randomized lossy
coding channel capacity:

\begin{eqnarray}
  \liminf_{n\rightarrow \infty} \frac{1}{n}I(\rva^{nR}; \rvb^n)\geq
\tilde R_p=\sup_{\lim\limits_{ n\rightarrow \infty}e_{p,n}(\tilde
R)=0}\{\tilde R\}\label{eqn.lowerbound_I(a,b)}
\end{eqnarray}

\end{lemma}

\proof: to show~\ref{eqn.lowerbound_I(a,b)}, from the definition of
$\tilde R_p$, we know that it is enough to show that for all $\tilde
R$, such that $\lim\limits_{ n\rightarrow \infty}e_{p,n}(\tilde
R)=0$:
$$\liminf_{n\rightarrow \infty} \frac{1}{n}I(\rva^{nR}; \rvb^n)\geq
\tilde R .$$

First we take a new perspective of the Bernoulli sequence $\rvb^n$.
Instead of letting $\rvb^n$ be i.i.d generated from the Bernoulli
$p$ random process, we first generate two auxiliary random variables
$\rvC_p$ and $\rvm$ and then the $\rvb^n$ is a function of the two
auxiliary random variables in a way such that $\rvb^n$ is an i.i.d
Bernoulli $p$ sequence.

We first generate a codebook random variable $\rvC_p$ according to
the  distribution described in~(\ref{eqn.randomized_distribution}),
where the code book $\rvC_p=C=(c_1,...,c_{2^{n\tilde R}})$ with the
following probability:
\begin{eqnarray}
\Pr(\rvC_p=C)=\prod_{m=1}^{2^{n\tilde R}}
p^{1(c_m)}(1-p)^{n-1(c_m)}\nonumber.
\end{eqnarray}
Then we pick the message random variable $\rvm$ according that is
uniform on $\{1,2,..., 2^{n\tilde R}\}$. Finally we let the binary
sequence $\rvb^n$ be a function of $\rvC_p$ and $\rvm$, such that
for $\rvC_p=C=(c_1,...,c_{2^{n\tilde R}})$ and $\rvm=m$, $b^n=c_m$.
It is easy to see that $\rvb^n$ chosen this way have the following
distribution:
\begin{eqnarray}
\Pr(\rvb^n=b^n)=p^{1(b^n)}(1-p)^{n-1(b^n)}\nonumber.
\end{eqnarray}
So we have the following Markov Chain:
\begin{eqnarray}
(\rvC_p, \rvm)\rightarrow \rvb^n
\rightarrow\rva^{nR}\label{eqn.Markovchain}
\end{eqnarray}
So from the data processing lemma and the chain rule for mutual
information, we know that:
\begin{eqnarray}
I(\rva^{nR}; \rvb^n)&\geq& I(\rva^{nR}; \rvC_p, \rvm )\nonumber\\
&=& I(\rva^{nR};  \rvm |\rvC_p) +  I(\rva^{nR};  \rvC_p)\nonumber\\
&\geq& I(\rva^{nR};  \rvm |\rvC_p)\label{eqn.lowerboundingIAB}
\end{eqnarray}
where the last inequality follows that mutual information is always
non-negative. Now the overall error probability is, as defined
in~(\ref{eqn.error_randomizedcode}):
\begin{eqnarray}
e_{p,n}(\tilde R)= \sum_{C\in \mathcal B_n ^{2^{n\tilde
R}}}\Pr(\rvC_p=C)\left(\Pr(\rvm \neq \hat
\rvm(\rva^{nR})|\rvC_p=C)\right)
\end{eqnarray}
where $\Pr(\rvC_p=C) (\Pr(\rvm \neq \hat \rvm(\rva^{nR})|\rvC_p=C))$
is the decoding error when the code book $C$ is chosen. Hence this
is a standard communication problem that we can use the technique
detailed in Chapter 8.9~\cite{Cover} to lower bound the mutual
information $I(\rva^{nR}; \rvb^n)$ by the rate $\tilde R$ that a
reliable communication is possible. Notice that if the codebook $C$
is chosen, we have the following Markov Chain:
\begin{eqnarray}
\rvm\rightarrow \rvb^n \rightarrow\rva^{nR}\rightarrow \hat
\rvm\label{eqn.Markovchain1},
\end{eqnarray}
more specifically $\rvb^n$ is a deterministic function of $\rvm$,
$\hat \rvm$ is a deterministic function of $\rva^{nR}$. So we can
apply Fano's inequality (Theorem 2.11.1~\cite{Cover} for any fixed
codebook $C$:
\begin{eqnarray}
H(\rvm|\rva^{nR}, \rvC_p=C) \leq 1+\Pr(\rvm \neq \hat
\rvm(\rva^{nR})|\rvC_p=C) n\tilde R
\end{eqnarray}
Now, from the standard information theoretic equalities:
\begin{eqnarray}
n\tilde R& = &H(\rvm)\nonumber\\
&=& H(\rvm|\rvC_p=C)\nonumber\\
&=&
H(\rvm|\rva^{nR},\rvC_p=C)+I(\rva^{nR};  \rvm |\rvC_p=C)\nonumber\\
&\leq &1+\Pr(\rvm \neq \hat \rvm(\rva^{nR})|\rvC_p=C) n\tilde R
+I(\rva^{nR};  \rvm |\rvC_p=C)\nonumber
\end{eqnarray}
Multiply both sides by $\Pr(\rvC_p=C)$ and sum over all
$C\in\mathcal B_n ^{2^{n\tilde R}}$, we have:

\begin{eqnarray}
n\tilde R &\leq &1+ \sum_{C\in \mathcal B_n ^{2^{n\tilde
R}}}\Pr(\rvC_p=C)\left(\Pr(\rvm \neq \hat \rvm(\rva^{nR})|\rvC_p=C)
n\tilde R +I(\rva^{nR};  \rvm |\rvC_p=C)\right)\nonumber\\
&=&1+n\tilde R\times e_{p,n}(\tilde R)+ I(\rva^{nR};  \rvm
|\rvC_p)\label{eqn.finalI(ab)}
\end{eqnarray}
Finally, substitute~(\ref{eqn.lowerboundingIAB})
into~(\ref{eqn.finalI(ab)}), we have:
\begin{eqnarray}
I(\rva^{nR}; \rvb^n)\geq I(\rva^{nR};  \rvm |\rvC_p)\geq  n\tilde R
-1-n\tilde R \times e_{p,n}(\tilde R)\nonumber
\end{eqnarray}
So, if the randomized lossy coding capacity is above $\tilde R$,
i.e. $\lim\limits_{n\rightarrow \infty}e_{p,n}(\tilde R)=0$, then
\begin{eqnarray}
  \liminf_{n\rightarrow \infty} \frac{1}{n}I(\rva^{nR}; \rvb^n)\geq
\tilde R\nonumber
\end{eqnarray}
 \hfill $\square$

\section{Randomized Channel capacity of a lossy compressor, a lower
bound}\label{sec.lowerboundingchannel}

In the previous section, we showed the relation between the mutual
information $I(\rva^{nR}, \rvb^n)$ is lower bounded by the
randomized lossy coding capacity if the input codewords look like an
i.i.d Bernoulli $p$ sequence.  What was missing in the previous
section is a lower bound on the randomized capacity. In this section
we  study the capacity, in particular the lower  bound on the
capacity. Notice that the encoder is using a randomized code book
according to the distribution
in~(\ref{eqn.randomized_distribution}). We only need to design the
decoder $G_n$ in Figure~\ref{fig.BG_Lossycoding_channel_coding}. If
we could show  that for some $\tilde R$, the average error
probability $e_{p,n}(\tilde R)$ goes to zero as $n$ goes to
infinity, then whatever the $\tilde R$ is, it is a lower bound on
the randomized lossy coding capacity $\tilde R_p$. We give a lower
bound on $\tilde R_p$. As will be clear soon from our derivation of
the lower bound, this bound is not tight. However, this is our first
effort to derive a non-trivial lower bound to the rate distortion
function $R(D,p)$.

\vspace{0.1in}
\begin{lemma}{A lower  bound on the randomized lossy coding capacity:}\label{lemma.lossycoding_capacity}

\begin{eqnarray}
\tilde R_p \geq \underline{ \tilde R}=\max_{L \geq 0}\{\min_{U\geq
L, r\in[0,1-p]: T_1(L,U,r)\leq D } h(L,U,r)  \}\nonumber
\end{eqnarray}

\begin{eqnarray}
\mbox{where   } h(L,U,r)=\{
\begin{array}{ccc}\label{eqn.h(L,U,r)}
 (p\times\Pr(|\rvs|> U)+r)D(\frac{p\times\Pr(|\rvs|> U)}{p\times\Pr(|\rvs|> U)+r}\|p) &\mbox{,  if }\frac{p\times\Pr(|\rvs|> U)}{p\times\Pr(|\rvs|> U)+r} \geq p &\\
0   &\mbox{,  if } \frac{p\times\Pr(|\rvs|> U)}{p\times\Pr(|\rvs|>
U)+r}< p&
\end{array}
\end{eqnarray}

\begin{eqnarray}
\rvs \mbox{ in~(\ref{eqn.h(L,U,r)}) is  Gaussian } N(0,1) \mbox{ and
}T_1(L,U,r)=r L^2 + 2p \int^{U}_{L} (s-L)^2
\frac{1}{\sqrt{2\pi}}e^{-\frac{ \svs^2}{2}}d s\nonumber
\end{eqnarray}

 Or equivalently, for all $\tilde R\leq \underline{ \tilde R}$, the
 decoding error defined in~(\ref{eqn.error_randomizedcode}) for the randomized coding scheme converges to zero
 as $n$ goes to infinity:
\begin{eqnarray}
\lim_{n\rightarrow \infty} e_{p,n}(\tilde R) =0 \nonumber
\end{eqnarray}
\end{lemma}
 \vspace{0.1in}

\proof we first describe the decoder $G_n$.  The codebook $C$ is
chosen, i.e. $\rvC_p=C$. As shown in
Figure~\ref{fig.BG_Lossycoding_channel_coding}, if a message $\rvm$
is to be sent, where $\rvm\in\{1,2,..., 2^{n\tilde R}\}$ with equal
probability, the binary output to the channel encoder $F_n$ is
$c_\rvm$. After the modulation of the Gaussian sequence $\rvs^n$ and
the lossy source coding encoder $f_n$, the channel decoder $G_n$
receives $\rva^{nR}$. The first step of $G_n$ is to run the lossy
source decoder $g_n$ and get the lossy estimate of
$\rvx^n=c_\rvm\times \rvs^n$, $\hat \rvx^n=g_n(\rva^{nR})$. The
second step of $G_n$ is to estimate $\rvm$ from $\hat \rvx^n$. We
pick the code word with the most entries' absolute value above the
positive real number $L$:
\begin{eqnarray}
\hat \rvm(\sva^{nR}(c_1\times \rvs^n)) =\hat \rvm(\hat \rvx^n)=
\argmax_{i}\sum_{k=1}^{n} 1(|c_i(k)\hat \rvx_k)| \geq L)
\label{eqn.channeldecoding}
\end{eqnarray}
where $c_i\in \{0,1\}^n$ is the codeword for message $i$ in the
chosen codebook $C$ and $c_i(k)\in\{0,1\}$ is the $k$-th entry of
the codeword $c_i$. Now we analyze the average error probability of
the above coding system over all codebooks according the the
codebook distribution in~(\ref{eqn.randomized_distribution}) and the
over all Gaussian sequence $\rvs^n$. The average error probability
is hence as shown in~(\ref{eqn.error_randomizedcode}):
\begin{eqnarray}
e_{p,n}(\tilde R)&= &\sum_{C\in \mathcal B_n ^{2^{n\tilde
R}}}\Pr(\rvC_p=C)\left(\frac{1}{2^{n\tilde R}}\sum_{m=1}^{2^{n\tilde
R}}\Pr^{\rvs}(m\neq \hat m (\sva^{nR}(c_m\times
\rvs^n)))\right)\nonumber\\
&=& \sum_{C\in \mathcal B_n ^{2^{n\tilde
R}}}\Pr(\rvC_p=C)\left(\Pr^{\rvs}(1\neq \hat m (\sva^{nR}(c_1\times
\rvs^n)))\right) \label{eqn.error_randomizedcode0}\\
&=& \Pr^{\rvC_p,\rvs}(1\neq \hat m (\sva^{nR}(c_1\times
\rvs^n)))\label{eqn.error_randomizedcode1}
\end{eqnarray}
where~(\ref{eqn.error_randomizedcode0}) follows the symmetry of the
system.

 We
decompose~(\ref{eqn.error_randomizedcode1}) into four parts.  We
sketch the partitions then give a detailed analysis.
 \begin{enumerate}
 \item The atypical behavior of codeword $\rvc_1$. The typicality is
 defined in the usual way~\cite{Cover} for finite discrete random
 sequences. The concentration theorem is well established in the
 literature.

 \item The atypical behavior of  $\widetilde{\rvs}^{1(\rvc_1)}$
 while
 $\rvc_1$ is typical, where $\widetilde \rvs^{1(\rvc_1)} $ is the non-zero subsequence of $\rvx^n=\rvc_1\times \rvs^n$
 where
 $\widetilde{\rvs}_1=\rvs_{i_1},...,\widetilde{\rvs}_{1(\rvc_1)}=\rvs_{i_{1(\rvc_1)}}$, where $i_1,...,
 i_{1(\rvc_1)}$ are the non-zero locations of $\rvc_1$. The typicality for a Gaussian $N(0,1)$ sequence is defined in
 Appendix~\ref{sec.strong_typical_gaussian}. We prove the
 concentration result in Lemma~\ref{lemma.concentrationGaussian}.

 \item The atypical behavior of the lossy source coding while both $\rvc_1$ and $\rvs_{i_1},...,\rvs_{i_{1(\rvc_1)}}$
are typical. i.e. the distortion of the Bernoulli-Gaussian sequence
$d(\rvc_1\times\rvs^n, \hat \rvx^n)= d(\rvx^n, \hat \rvx^n)> D$, the
concentration of the typical behavior of the lossy source coding is
established in~(\ref{eqn.codingpair_BG}) for good lossy coders.

 \item The probability that there exists a message $\underline{m}$
 that has a higher score than message $1$ according to the decoding
 rule in~(\ref{eqn.channeldecoding}) while everything else (the codeword for message
 $1$, $\rvc_1$, the subsequence
 $\rvs_{i_1},...,\rvs_{i_{1(\rvc_1)}}$, and the distortion $d(\rvc_1\times\rvs^n, \hat
 \rvx^n)$ are typical. We bound this error by a union bound
 argument.\\

 \end{enumerate}

\textbf{The first part} is the atypicality of the codeword for
message $1$, $\rvc_1$, the second part is the error probability for
$\rvc_1\in \mathcal B^n_\epsilon$,  where
\begin{eqnarray}
B^n_{\epsilon}\triangleq\{b^n\in \{0,1\}^n: |\frac{1(b^n)}{n}-p|\leq
\epsilon\}.\nonumber
\end{eqnarray}
Under the codebook probability $\rvC_p$, all $\rvc_i$'s are binary
sequences of length-$n$ with distribution  such that for all $b^n\in
\{0,1\}^n$:
\begin{eqnarray}
\Pr^{\rvC_p}(\rvc_i=b^n)= p^{1(b^n)}(1-p) ^{n-1(b^n)}, \ \
i=1,2,...2^{n\tilde R}.
\end{eqnarray}
so we obviously have~\cite{Cover}:
\begin{eqnarray}
\lim_{n\rightarrow \infty}\Pr^{\rvC_p} (\rvc_1\notin
B^n_{\epsilon})=0\label{eqn.typicalpart1}
 \end{eqnarray}
\vspace{0.1in}

\textbf{The second part} is the atypicality of the Gaussian
subsequence
 $\rvs_{i_1},...,\rvs_{i_{1(\rvc_1)}}$, where $i_1,...,
 i_{1(\rvc_1)}$ are the non-zero locations of $\rvc_1$, while
 $\rvc_1$ is typical, $\rvc_1\in B^n_{\epsilon})$. The typical
 Gaussian $N(0,1)$ set is defined as follows, first we have two
 definitions: for a real sequence $\svs^n$ and s.t. $-\infty\leq S\leq T\leq \infty$,
 the
$l$-th moment of entries in $\svs^n$ within interval  $[S, T]$ is
denoted by
$$n^l_{\svs^n}(S,T)=\frac{\sum_{i=1}^n 1(S<\svs_i<T)\svs_i^l}{n} .$$

Then the $\epsilon$-typical set for Gaussian $N(0,1)$ is defined as:

 \begin{eqnarray}
S_{\epsilon}(n)=\left\{\svs^n:
\max_{l=0,1,2}\left\{\sup_{S,T}\left|n^l_{\svs^n}(S,T)-\int_S^{T}\svs^l\frac{1}{\sqrt{2\pi}}e^{\frac{-s^2}{2}}ds\right|<\epsilon\right\}\right\}\nonumber
\end{eqnarray}

 We prove the concentration result in
 Lemma~\ref{lemma.concentrationGaussian} in Appendix~\ref{sec.strong_typical_gaussian}:
$ \lim\limits_{n\rightarrow \infty}\Pr\limits^{\rvs} (\rvs^n\notin
S_{\epsilon}(n))=0$. $\rvc_1$ and $\rvs^n$ are independent, and if
$\svc_1\in B_\epsilon^n$, then $1(\svc_1)\geq p(n-\epsilon)$, so if
$n$ goes to infinity, $1(\svc_1)$ goes to infinity too,  so
\begin{eqnarray}
\lim _{n\rightarrow \infty}\Pr ^{\rvC_p, \rvs} (\rvc_1\in
B_\epsilon^n, \widetilde{\rvs}^{1(\rvc_1)}\notin
S_{\epsilon}({1(\rvc_1)}))&\leq &\lim _{n\rightarrow \infty}\Pr
^{\rvC_p, \rvs} ( \widetilde{\rvs}^{1(\rvc_1)}\notin
S_{\epsilon}({1(\rvc_1)})|\rvc_1\in B_\epsilon^n)\nonumber\\
&=&0\label{eqn.typicalpart2}
\end{eqnarray}
where the first inequality follows that conditional probability is
bigger than joint probability.\\

\textbf{The third part} is the atypical behavior of the lossy coding
system. Following the definition of a good lossy source coder in the
strong sense in~(\ref{eqn.codingpair_BG}) and that
$\rvx^n=\rvc_1\times \rvs^n$, we have, for all $\delta_1>0$
\begin{eqnarray}
\lim_{n\rightarrow \infty} \Pr^\rvx\left(d(\rvx^n, \hat \rvx^n)\geq
D+\delta_1\right)= \lim_{n\rightarrow \infty}
\Pr^{\rvC_p,\rvs}\left(d(\rvc_1\times\rvs^n, \hat \rvx^n)\geq
D+\delta_1\right)=0\nonumber
\end{eqnarray}

This implies that:
\begin{eqnarray} \lim_{n\rightarrow \infty}
\Pr^{\rvC_p,\rvs}\left(\rvc_1\in B_\epsilon^n,
\widetilde{\rvs}^{1(\rvc_1)}\in S_{\epsilon}({1(\rvc_1)}),
d(\rvc_1\times\rvs^n, \hat \rvx^n)\geq
D+\delta_1\right)=0\label{eqn.typicalpart3}
\end{eqnarray}
\vspace{0.1in}

\textbf{The fourth part} is when the code word $\rvc_1$, the
Gaussian subsequence $\widetilde \rvs^{1(\rvc_1)}$, and the
distortion $d(\rvc_1\times\rvs^n, \hat \rvx^n)$ are all typical, the
decoding error for the channel decoder following the decoding rule
in~(\ref{eqn.channeldecoding}).

The output of the lossy source coding decoder is $\hat\rvx^n =
g_n(\sva^{nR}(c_1\times \rvs^n))$, from the decoding rule
in~(\ref{eqn.channeldecoding}),  the estimate of the message $\hat m
(\sva^{nR}(c_1\times \rvs^n)))$ is not equal to the true message
$1$, if and only if there exists a message $\underline{m}\neq 1$,
such that
\begin{eqnarray}
\sum_{k=1}^{n} 1(|c_{\underline{m}}(k)\hat \svx_k)| \geq L) \geq
\sum_{k=1}^{n} 1(|c_1(k)\hat \svx_k)| \geq L) \label{eqn.errorevent}
\end{eqnarray}
Notice that the codebooks are symmetric to the messages, i.e. over
all the codebooks, the probability that the estimation of the
message $\hat \rvm =i$ is equal to the probability that $\hat \rvm
=j$ for all $i,j \in\{1,2,...,2^{n\tilde R}\}$ and $i\neq 1$, $j\neq
1$. So we can union bound the  decoding error probability of the
event shown in~(\ref{eqn.errorevent}) as follows:
\begin{eqnarray}
 \Pr^{\rvs,\rvC_p}(1\neq \hat m (\sva^{nR}(\rvc_1\times \rvs^n)))
&\leq &   2^{n\tilde R} \Pr^{\rvs,\rvC_p}\left(\sum_{k=1}^{n}
1(|\rvc_{2}(k)\hat \svx_k | \geq L) \geq \sum_{k=1}^{n}
1(|\rvc_1(k)\hat \svx_k | \geq L)\right)
\label{eqn.decodingerror_uv}
\end{eqnarray}
where the probability is calculated over all possible codebooks over
the measure $\rvC_p$ and the Gaussian sequences $\rvs^n$. First, for
a codeword $c_1$, and the lossy coding estimate of $c_1\times
\rvs^n$, $\hat \svx^n$, denote by $u$ and $v$ the number of entries
of the estimate $\hat \svx_k$ with absolute value above $L$ where
$c_1(k)$ is $1$ and $0$ respectively:
\begin{eqnarray}
u &= & \sum_{k=1}^{n} 1(|c_1(k)\hat \svx_k)| \geq L)\nonumber\\
v&= & \sum_{k=1}^{n} 1( |\hat \svx_k)| \geq L \mbox{ and }
c_1(k)=0)\label{eqn.defintionUV}.
\end{eqnarray}
With $u$ and $v$ fixed( here we fix the codeword $c_1$, the sequence
$\svs^n$ and the estimate $\hat \svx^n$), we union bound the
probability of the following event that there exists a message
$\underline{m}\neq 1$, such that~(\ref{eqn.errorevent}) is true:
\begin{eqnarray}
 \Pr^{\rvs,\rvC_p}(1\neq \hat m (\sva^{nR}(\rvc_1\times \rvs^n))|\rvc_1=\svc_1, \rvs^n=\svs^n)&\leq&
    2^{n\tilde R}\Pr^{\rvC_p}\left(\sum_{k=1}^{n} 1(|
\rvc_{2}(k)\hat \svx_k| \geq L) \geq \sum_{k=1}^{n} 1(|c_1(k)\hat
\svx_k| \geq L) \right)\nonumber\\
&= &   2^{n\tilde R}\Pr^{\rvC_p}\left(\sum_{k=1}^{n} 1(|
\rvc_{2}(k)\hat \svx_k| \geq L) \geq u \right) \label{eqn.boundingC2.1}\\
&= &   2^{n\tilde R}
\sum_{l=u}^{u+v}\nchoosek{u+v}{l}p^l(1-p)^{u+v-l}
\label{eqn.boundingC2.2} \\
&\leq&  2^{n\tilde R}\times n \max_{l: u\leq l\leq u+v}\{
\nchoosek{u+v}{l}p^l(1-p)^{u+v-l}\}\label{eqn.boundingC2.2.1} \\
&\leq &   2^{n\tilde R}\times  n 2^{-  (u+v)\min\limits_{l: u\leq
l\leq u+v} D(\frac{l}{u+v}\|
p)}\label{eqn.boundingC2.3}\\
&=&  2^{n\tilde R}\times \{
\begin{array}{ccc}
 n &\mbox{,  if } \frac{u}{u+v}\leq p  &\label{eqn.boundingC2.4}\\
 n 2^{-(u+v) D(\frac{u}{u+v}\| p )}   &\mbox{,  if } \frac{u}{u+v}> p&
\end{array}
\end{eqnarray}

(\ref{eqn.boundingC2.1}) follows the definition of $u$.
(\ref{eqn.boundingC2.2}) follows that $\rvc_2\in \{0,1\}^n$ is an
i.i.d. Bernoulli $p$ sequence. (\ref{eqn.boundingC2.2.1}) is because
$u+v\leq n$. (\ref{eqn.boundingC2.3}) and~(\ref{eqn.boundingC2.4})
follows basic information theoretic inequalities~\cite{csiszar}.
From Lemma~\ref{lemma.property_of_EE} in
Appendix~\ref{sec.property_of_EE}, we know that the $(u+v)
D(\frac{u}{u+v}\| p )$ is monotonically increasing with $u$ and
monotonically decreasing with $v$. $\frac{u}{u+v}$ is also
monotonically increasing with $u$ and monotonically decreasing with
$v$, so the expression in~(\ref{eqn.boundingC2.4}) is monotonically
\textbf{decreasing} with $u$ and monotonically \textbf{increasing}
with $v$.

(\ref{eqn.boundingC2.4})  is true for all codeword $\svc_1$ and
sequence $\tilde \svs^{1(\svc_1)}$, typical or not. So it is also
true for all those  $\svc_1\in B_\epsilon^n$, $\tilde
\svs^{1(\svc_1)}\in S_{\epsilon}({1(\rvc_1)})$ and
$d(\svc_1\times\svs^n, \hat \svx^n)\leq  D+\delta_1$ in this case,
we can give a feasible region for $u $ and $v$, i.e.  then give a
bound on (\ref{eqn.boundingC2.4}). We further investigate the
distortion for the said typical sequences:
\begin{eqnarray}
 n(D+\delta_1)&\geq& n d(\svc_1\times\svs^n, \hat \svx^n)\nonumber\\
 &=&\sum_{k=1}^n (\svc_1(k)\svs_k-\hat \svx_k)^2\nonumber\\
&=&\sum_{k: \svc_1(k)=1} (\svc_1(k)\svs_k-\hat \svx_k)^2
+ \sum_{k: \svc_1(k)=0} \hat \svx_k^2\nonumber\\
&=&\sum_{k: \svc_1(k)=1} (\svc_1(k)\svs_k-\hat \svx_k)^2
+ \sum_{k: \svc_1(k)=0, \svx_k \geq L} \hat \svx_k^2+ \sum_{k: \svc_1(k)=0, \svx_k < L} \hat \svx_k^2\nonumber\\
&\geq&\sum_{k: \svc_1(k)=1} (\svc_1(k)\svs_k-\hat \svx_k)^2 +
vL^2\label{eqn.bounding_vL}
\end{eqnarray}

where~(\ref{eqn.bounding_vL}) follows the definition of $v$. Notice
that by definition $x_k=\svc_1(k)\svs_k$, so $\svx_k>0$ implies that
$\svc_1(k)=1$, the first term of~(\ref{eqn.bounding_vL}) is:

\begin{eqnarray}
\sum_{k: \svc_1(k)=1} (x_k -\hat \svx_k)^2 &\geq &
 \sum_{k:  |x_k|\geq  L\geq |\hat \svx_k |} (x_k -\hat \svx_k)^2\nonumber\\
&\geq &
 \sum_{k:|x_k|\geq  L\geq |\hat \svx_k | } (|x_k| -L
 )^2\nonumber
\end{eqnarray}
We rewrite~(\ref{eqn.bounding_vL}) as:
\begin{eqnarray}
 n(D+\delta_1) \geq
 \sum_{k:|x_k|\geq  L\geq |\hat \svx_k | } (|x_k| -L
 )^2+ vL^2 \label{eqn.bounding_vL1}
\end{eqnarray}
From the definition of $u$: we know that
$ u= \sum\limits_{k=1}^{n} 1(|c_1(k)\hat \svx_k)| \geq L)$ hence
\begin{eqnarray}
\sum\limits_{k=1}^{n} 1(|x_k|\geq  L\geq |\hat \svx_k |) & \geq &
\sum\limits_{k=1}^{n} 1(|x_k|>0)- \sum\limits_{k=1}^{n} 1(0< |\svx_k
|\leq L)- \sum\limits_{k=1}^{n}
1(|c_1(k)\hat \svx_k)| \geq L)\nonumber\\
&=& \sum\limits_{k=1}^{n} 1(|x_k|>L) -u\nonumber\\
&\triangleq& n(|x_k|>L) -u \label{eqn.bounding_vL2}
\end{eqnarray}
Recall that $\tilde s_1,... \tilde s_{1(\svc_1)}$ are the none-zero
entries of $\svx^n$, without out loss of generality, let $|\tilde
s_1|,... |\tilde s_{n(|x_k|>L)-u}|$ be the smallest $ n(|x_k|>L)-u$
many $|\svx_k|$'s that are larger than $L$, without loss of
generality let $|\tilde s_1|\geq  .... \geq |\tilde
s_{n(|x_k|>L)-u}|\geq L $. Then
substituting~(\ref{eqn.bounding_vL2}) into~(\ref{eqn.bounding_vL1})
 and denote by $\tilde U =|\tilde\svs_1|$, we have:
\begin{eqnarray}
 n(D+\delta_1) &\geq&
 \sum_{j=1 }^{n(|x_k|>L)-u } (|\tilde \svs_j|  -L
 )^2+ vL^2\nonumber\\
&=&
 \sum_{j: L< |\tilde \svs_j|\leq \tilde U } (|\tilde \svs_j | -L
 )^2+ vL^2\label{eqn.bounding_vL3.1}\\
&=&
 \sum_{j: L< |\tilde\svs_j|\leq \tilde U }  (|\tilde  \svs_j| ^2 -2L  |\tilde
 \svs_j| +L^2
 )+ vL^2\nonumber\\
&\geq & 2\times 1(\svc_1)\left(\int^{\tilde U}_{L} (s-L)^2
\frac{1}{\sqrt{2\pi}}e^{-\frac{ \svs^2}{2}}d s
- \epsilon (1+L)^2\right)+ vL^2\label{eqn.bounding_vL3.2}\\
&\geq & 2\times n(p-\epsilon)\left(\int^{\tilde U}_{L} (s-L)^2
\frac{1}{\sqrt{2\pi}}e^{-\frac{ \svs^2}{2}}d s - \epsilon
(1+L)^2\right)+ vL^2
  \label{eqn.bounding_vL3.3}\\
&\geq & n \left( 2 p \int^{\tilde U}_{L} (s-L)^2
\frac{1}{\sqrt{2\pi}}e^{-\frac{ \svs^2}{2}}d s
+\frac{v}{n}L^2\right)- n \epsilon K_1(p, L)
  \label{eqn.bounding_vL3}
\end{eqnarray}
 (\ref{eqn.bounding_vL3.1})
follows the definition of $\tilde s^{1(\svc_1)}$,
(\ref{eqn.bounding_vL3.2}) is true because $\tilde s^{1(\svc_1)}\in
S_\epsilon(1(\svc_1))$ is $\epsilon$-typical Gaussian $N(0,1)$.
(\ref{eqn.bounding_vL3.3}) is true because $\svc_1\in B_\epsilon
^n$. Finally in (\ref{eqn.bounding_vL3}),  $K_1(p,L)$ is a finite
function of $p$ and $L$, we do not need $\tilde U$ in the picture
because we can replace $\tilde U$ with $\infty$ when bounding the
the residue. We rewrite~(\ref{eqn.bounding_vL3}) as:
\begin{eqnarray}
 2 p \int^{\tilde U}_{L} (s-L)^2 \frac{1}{\sqrt{2\pi}}e^{-\frac{
\svs^2}{2}}d s +\frac{v}{n}L^2 \leq D+\delta_1 +  \epsilon K_1(p, L)
  \label{eqn.bounding_vL4}
\end{eqnarray}

Meanwhile, because $\tilde U=|\tilde s_1|\geq... \geq |\tilde
s_{n(|x_k|>L)-u}|\geq L$ are the smallest $ n(|x_k|>L)-u$ many
$|\svx_k|$'s that are larger than $L$, $\tilde \svs ^{1(\svc_1)}$ is
a $\epsilon$-typical Gaussian sequence, so $n( |x_k|>L )-u\leq
1(\svc_1)(\Pr(L<|\rvs|<\tilde U )+\epsilon)$, hence:
\begin{eqnarray}
u&>& n( |x_k|>L )- 1(\svc_1)(\Pr(L<|\rvs|<\tilde U
)+\epsilon)\nonumber\\
&\geq& n
(p-\epsilon)(\Pr(|\rvs|>L)-\epsilon)-n(p+\epsilon)(\Pr(L<|\rvs|<\tilde
U
)+\epsilon)\nonumber\\
&=&np\Pr(|\rvs|>\tilde U)- n\epsilon K_2(p,
L)\label{eqn.bounding_vL5}
\end{eqnarray}
The above analysis are true for all $\delta_1$ and $\epsilon$, we
let both be small, we have
\begin{eqnarray}
&&u\geq n\big(p\Pr(|\rvs|> U)-\epsilon_2\big)\label{eqn.bounding_vL6}\\
&&\mbox{ s.t.: } 2 p \int^{ U}_{L} (s-L)^2
\frac{1}{\sqrt{2\pi}}e^{-\frac{ \svs^2}{2}}d s +\frac{v}{n}L^2 \leq
D\label{eqn.bounding_vL6.1}
\end{eqnarray}
where $\lim\limits_{\delta, \epsilon\rightarrow 0} \epsilon_2=0$,
this is true because  for any $\tilde U$ that
satisfies~(\ref{eqn.bounding_vL4}), it either also satisfies the
more stringent constraint in~(\ref{eqn.bounding_vL6.1}) or the gap
between $\tilde U$ and the biggest $U$ that
satisfies~(\ref{eqn.bounding_vL6.1}) is small when $\delta_1$ and
$\epsilon$ are small.  Then~(\ref{eqn.bounding_vL6.1}) follows the
continuity of $\Pr(|\rvs|>U)$ in $U$.

 \vspace{0.1in}

Notice that~(\ref{eqn.boundingC2.4}) holds for all codeword $\svc_1$
and $\svs^n$, in particular it is true for the typical ones,
$\svc_1\in B_\epsilon^n$ and $\tilde \svs^{1(\svc_1)}\in
S_\epsilon(1(\svc_1))$ and $d(\svc_1\times\svs^n, \hat \svx^n)\leq
D+\delta_1$, also~(\ref{eqn.boundingC2.4})  is monotonically
decreasing with $u$, with~(\ref{eqn.bounding_vL6}) and let
$r=\frac{v}{n}$, recall the definition of $v$
in~(\ref{eqn.defintionUV}), for $\svc_1\in B^\epsilon_n$, $v\leq n-
n(p-\epsilon)= n(1-p+\epsilon)$ or equivalently $r\in [0, 1-p]$, we
rewrite~(\ref{eqn.boundingC2.4}):
\begin{eqnarray}
&& \Pr^{\rvs,\rvC_p}(1\neq \hat m (\sva^{nR}(\rvc_1\times
\rvs^n))|\rvc_1=\svc_1\in B_\epsilon^n,
  \rvs^n=\svs^n\in
S_\epsilon(1(\svc_1)), d(\svc_1\times\svs^n, \hat \svx^n)\leq  D+\delta_1)\nonumber\\
&\leq&
 2^{n\tilde R}\times \{
\begin{array}{ccc}
 n &\mbox{,  if } \frac{u}{u+v}\leq p  &\nonumber\\
 n 2^{-(u+v) D(\frac{u}{u+v}\| p )}   &\mbox{,  if } \frac{u}{u+v}> p&
\end{array}\nonumber\\
&\leq&
 2^{n\tilde R}\times \{
\begin{array}{ccc}
 n &\mbox{,  if } \frac{\Pr(|\rvs|>U)}{\Pr(|\rvs|>U)+r}\leq p  &\label{eqn.final_boundRTILDE}\\
 n 2^{-n\left((\Pr(|\rvs|>U)+r) D( \frac{\Pr(|\rvs|>U)}{\Pr(|\rvs|>U)+r}\| p )-\epsilon_3\right)}   &\mbox{,  if }  \frac{\Pr(|\rvs|>U)}{\Pr(|\rvs|>U)+r}> p&
\end{array}
\end{eqnarray}
with~(\ref{eqn.bounding_vL6.1}) being satisfied,  where
$\lim\limits_{\epsilon_2\rightarrow 0}\epsilon_3=0$ because the
exponent in~(\ref{eqn.final_boundRTILDE}) is continuous in $ u $, we
know that $\lim\limits_{\delta, \epsilon\rightarrow 0}
\epsilon_2=0$, so $\lim\limits_{\delta, \epsilon\rightarrow 0}
\epsilon_3=0$ as well.

Notice that the coding system can pick arbitrary $L$, it picks the
best possible $L$, we have, if
\begin{eqnarray}
\tilde R< \underline{ \tilde R}=\max_{L \geq 0}\{\min_{U\geq L,
r\in[0,1-p]: T_1(L,U,r)\leq D } h(L,U,r)  \}\nonumber
\end{eqnarray}

\begin{eqnarray}
\mbox{where   } h(L,U,r)=\{
\begin{array}{ccc}\nonumber
 (p\times\Pr(|\rvs|> U)+r)D(\frac{p\times\Pr(|\rvs|> U)}{p\times\Pr(|\rvs|> U)+r}\|p) &\mbox{,  if }\frac{p\times\Pr(|\rvs|> U)}{p\times\Pr(|\rvs|> U)+r} \geq p &\\
0   &\mbox{,  if } \frac{p\times\Pr(|\rvs|> U)}{p\times\Pr(|\rvs|>
U)+r}< p&
\end{array}
\end{eqnarray}
then
\begin{eqnarray}
\lim_{n\rightarrow \infty} \Pr^{\rvs,\rvC_p}(1\neq \hat m
(\sva^{nR}(\rvc_1\times \rvs^n))|\rvc_1=\svc_1\in B_\epsilon^n,
  \rvs^n=\svs^n\in
S_\epsilon(1(\svc_1)), d(\svc_1\times\svs^n, \hat \svx^n)\leq
D+\delta_1)=0\label{eqn.typicalpart4.1}
\end{eqnarray}
The above inequality is true for  all those $\svc_1\in
B_\epsilon^n$, $\tilde \svs^{1(\svc_1)}\in
S_{\epsilon}({1(\rvc_1)})$ and $d(\svc_1\times\svs^n, \hat
\svx^n)\leq  D+\delta_1$ , so
\begin{eqnarray}
\lim_{n\rightarrow \infty} \Pr^{\rvs,\rvC_p}(1\neq \hat m
(\sva^{nR}(\rvc_1\times \rvs^n)), \rvc_1 \in B_\epsilon^n,
  \rvs^n \in
S_\epsilon(1(\svc_1)), d(\rvc_1\times\rvs^n, \hat \rvx^n)\leq
D+\delta_1)=0\label{eqn.typicalpart4}
\end{eqnarray}

\vspace{0.1in}

Finally we can upper bound the overall error probability of the
randomized coding scheme. The decoding error $e_{p,n}(\tilde R)$ is
defined in~(\ref{eqn.error_randomizedcode}) which is equivalent
to~(\ref{eqn.error_randomizedcode0}) because of the symmetry. We
decompose the error event into $4$ atypical events as illustrated at
the beginning of the proof. For any $\tilde R < \underline {\tilde
R}$,
\begin{eqnarray}
e_{p,n}(\tilde R)& =& \Pr^{\rvC_p,\rvs}(1\neq \hat m
(\sva^{nR}(c_1\times \rvs^n)))\label{eqn.final_decompose1}\\
&\leq& \Pr^{\rvC_p} (\rvc_1\notin B^n_{\epsilon})\nonumber\\
&& + \Pr ^{\rvC_p, \rvs} (\rvc_1\in B_\epsilon^n,
\widetilde{\rvs}^{1(\rvc_1)}\notin
S_{\epsilon}({1(\rvc_1)}))\nonumber\\
&& +\Pr^{\rvC_p,\rvs}\left(\rvc_1\in B_\epsilon^n,
\widetilde{\rvs}^{1(\rvc_1)}\in S_{\epsilon}({1(\rvc_1)}),
d(\rvc_1\times\rvs^n, \hat \rvx^n)> D+\delta_1\right)\nonumber\\
&& +\Pr^{\rvs,\rvC_p}(1\neq \hat m (\sva^{nR}(\rvc_1\times \rvs^n)),
\rvc_1 \in B_\epsilon^n,
  \rvs^n \in
S_\epsilon(1(\svc_1)), d(\rvc_1\times\rvs^n, \hat \rvx^n)\leq
D+\delta_1)\label{eqn.final_decompose2}
\end{eqnarray}
where~(\ref{eqn.final_decompose1})
follows~(\ref{eqn.error_randomizedcode0}). The asymptotic behaviors
of the four terms in~(\ref{eqn.final_decompose1}) are shown
in~(\ref{eqn.typicalpart1}),~(\ref{eqn.typicalpart2}),~(\ref{eqn.typicalpart3})
and~(\ref{eqn.typicalpart4}) respectively. $\delta_1$ can be
arbitrarily small, so we can finally claim that: for a good lossy
source coding system in the strong sense with distortion constraint
$D$, the randomized channel coding error converges to zero as $n$
goes to infinity:
\begin{eqnarray}
\lim_{n\rightarrow \infty} e_{p,n}(\tilde R) =0 \nonumber
\end{eqnarray}

This concludes the proof of Lemma~\ref{lemma.lossycoding_capacity}.
 \hfill $\square$

\section{Discussions and Numerical Result}\label{sec.discussions_num}
Now we have two upper bounds and two lower bounds on the rate
distortion function $R(D,p)$. We reiterate the bounds,
\begin{eqnarray}
&&R(D,p)\leq H(p) +  p R({D}, N(0,p))\label{eqn.bound1}\\
&& R(D,p)\leq    R(D, N(0,p))\label{eqn.bound2}\\
&& R(D,p)\geq    p R({D}, N(0,p))\label{eqn.bound3}\\
&& R(D,p)\geq   p R({D}, N(0,p))+ \max_{L \geq 0}\{\min_{U\geq L,
r\in[0,1-p]: T_1(L,U,r)\leq D } h(L,U,r)  \}\triangleq p R({D}, N(0,p))+ R_{i}(D,p)\label{eqn.bound4}\\
&&\ \ \ \mbox{where   } h(L,U,r)=\{
\begin{array}{ccc}\label{eqn.strings_attached}
 (p\times\Pr(|\rvs|> U)+r)D(\frac{p\times\Pr(|\rvs|> U)}{p\times\Pr(|\rvs|> U)+r}\|p) &\mbox{,  if }\frac{p\times\Pr(|\rvs|> U)}{p\times\Pr(|\rvs|> U)+r} \geq p &\\
0   &\mbox{,  if } \frac{p\times\Pr(|\rvs|> U)}{p\times\Pr(|\rvs|>
U)+r}< p&
\end{array}\\
&& \ \ \ \rvs  \mbox{ is  Gaussian } N(0,1) \mbox{ and }T_1(L,U,r)=r
L^2 + 2p \int^{U}_{L} (s-L)^2 \frac{1}{\sqrt{2\pi}}e^{-\frac{
\svs^2}{2}}d s\nonumber
\end{eqnarray}

where $R(D, N(0,p))$ is the rate distortion function for zero mean
variance $p$ Gaussian random sequence with distortion constraint
$D$, $R(D, N(0,p))=\max\{0, \frac{1}{2}\log_2\frac{p}{D}\}$.
(\ref{eqn.bound1}), (\ref{eqn.bound2}) and~(\ref{eqn.bound3}) are
derived in Propositions~\ref{prop.upper1}, ~\ref{prop.upper2}
and~\ref{prop.lower_trivial} respectively, (\ref{eqn.bound4}) is the
main result in Theorem~\ref{THM.mainresult}.

\subsection{Properties of the improvement $R_{i}(D,p)$}

The improvement of our new lower bound, the second term $R_{i}(D,p)$
in~(\ref{eqn.bound4}),  has a game theoretic interpretation.  In a
two player zero sum game, the first player (the coding system)
chooses $L$, the second player (adversary) chooses $U$ and $r$ with
string attached in~(\ref{eqn.strings_attached}), the payoff to
player one is $h(U,L,r)$. First we argue that the improvement of our
lower bound, the second term $R_{i}(p,D)$ in~(\ref{eqn.bound4}), is
monotonically decreasing with $D$ and if for some $D$, the
improvement is zero.\vspace{0.1in}
\begin{corollary}{$R_{i}(D,p)$ is monotonically decreasing with
$D$}, i.e. for $D_1> D_2$, $ R_{i}(D_1,p)\leq  R_{i}(D_2,p)$
\end{corollary}
\vspace{0.1 in}

\proof $R_{i}(D,p)$ is of the form of
\begin{eqnarray*}
\max_{L \geq 0}\{\min_{U\geq L, r\in[0,1-p]: T_1(L,U,r)\leq D }
h(L,U,r)  \},
\end{eqnarray*}
so for all $L\geq 0$, if the pair $(U,r)$ is  feasible for $D_2$, it
is also feasible for $D_1$, hence the minimum of $h(L,U,r)$ for
$D_1$ is no bigger than that for $D_2$. \hfill $\square$
\vspace{0.1in}

 More importantly the improvement is within $[0,
 H(p)]$ in light of the upper bound in~(\ref{eqn.bound1}). In the
 low distortion regime, i.e. $\frac{D}{p}\ll 1$. We argue that the
 improvement $R_{i}(D,p)$ is close to $ p \log_2 \frac{1}{p}$.
\vspace{0.1in}
\begin{corollary}{Asymptotic behavior of $R_{i}(D,p)$ in the
low distortion regime }\label{cor.asy}, for any $p>0$
\begin{eqnarray*}
\lim_{D\rightarrow 0}R_{i}(D,p) =p\log_2\frac{1}{p}
\end{eqnarray*}

\end{corollary}
\vspace{0.1 in}

\proof  We only give a sketch of proof here.    The coding system
pick a positive $L\ll 1$, but $L^2 \gg D$, say $L =D^{0.3}$
 The distortion constraint on $T_1(L, U, r)$ implies that $D \geq
 rL^2$, hence
\begin{eqnarray}
r \leq \frac{D}{L^2}= D^{0.4}.\nonumber
\end{eqnarray}
So $r$ goes to zero as $D$ goes to zero. Similarly we argue that $U$
goes to zero as $D$ goes to zero. In light of the distortion
constraint and that $L$ is picked to be $D^{0.3}$, also the obvious
inequality that $ - 2 sL \geq -\frac{s^2}{4}-4L^2$ for all $s$ and
$L$:

\begin{eqnarray}
\frac{D}{2p}  \geq   \int^{U}_{L} (s-L)^2
\frac{1}{\sqrt{2\pi}}e^{-\frac{ \svs^2}{2}}d s  \geq   \int^{U}_{L}
( \frac{3s^2}{4}-3L^2) \frac{1}{\sqrt{2\pi}}e^{-\frac{ \svs^2}{2}}d
s
= \int^{U}_{D^{0.3}} ( \frac{3s^2}{4}-3D^{0.6} )
\frac{1}{\sqrt{2\pi}}e^{-\frac{ \svs^2}{2}}d s\nonumber
\end{eqnarray}
hence:
\begin{eqnarray}
\int^{U}_{D^{0.3}}  \frac{3s^2}{4} \frac{1}{\sqrt{2\pi}}e^{-\frac{
\svs^2}{2}}d s \leq \frac{D}{2p} + \int^{U}_{D^{0.3}}  3D^{0.6}
\frac{1}{\sqrt{2\pi}}e^{-\frac{ \svs^2}{2}}d s \leq \frac{D}{2p} +
3D^{0.6}\nonumber
\end{eqnarray}
take limit on both side when $D\rightarrow 0$, the right hand side
is $0$, the left hand side is zero if and only if $U\rightarrow 0$
as $D$ goes to zero. We just showed that if we pick  $L=D^{0.3}$ and
$D$ goes to zero, then both $U$ and $r$ goes to zero if the
distortion constraint be satisfied. This means that the in this
case:
\begin{eqnarray*}
\lim_{D\rightarrow 0}R_{i}(D,p) =\lim_{r, U \rightarrow 0}
(p\times\Pr(|\rvs|> U)+r)D(\frac{p\times\Pr(|\rvs|>
U)}{p\times\Pr(|\rvs|> U)+r}\|p)  =p D(1\| p)=p \log_2\frac{1}{p }
\end{eqnarray*}
\hfill $\square$

\vspace{0.1in}

A simple corollary of Corollary~\ref{cor.asy} is  as follows. For
small $p$,   the sparse signal studied in the compressive sensing
literature:
\begin{eqnarray*}
H(p)= p \log_2(\frac{1}{p})+ (1-p) \log_2(\frac{1}{1-p})=   p
\log_2(\frac{1}{p})+  \log_2 (e) p
\end{eqnarray*}
So the gap between the improved lower bound in~(\ref{eqn.bound4})
and the upper bound in~(\ref{eqn.bound1}) is at most $\log_2(e) p$
which is dominated by the improvement $p \log_2\frac{1}{p}$ for
small $p$.

\subsection{Numerical Results}

\begin{figure}[htp]
\begin{center}
\includegraphics[width=120mm]{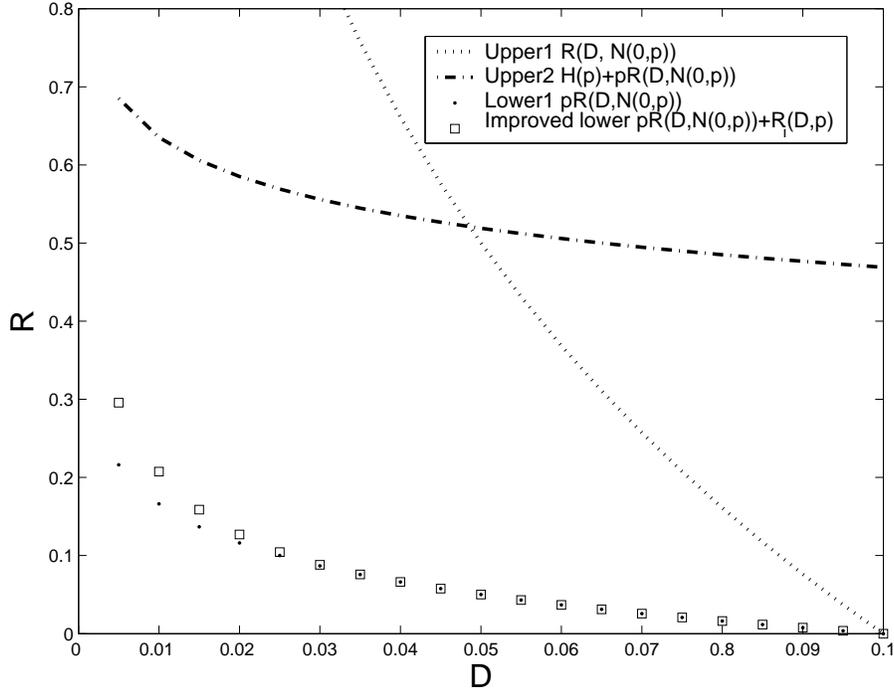}
\end{center} \caption[ ]{ Lower and upper bounds on $R(D,p)$ for
$p=0.1$ at  high distortion levels, the distortion $D$ runs from
$0.005$ to $0.1$}
    \label{fig.Numerical}
\end{figure}

 We plot the bounds in~(\ref{eqn.bound1})-~(\ref{eqn.bound4}) for $p=0.1$.
As shown in Figure~\ref{fig.Numerical}, the rate distortion function
$R(D, p) $ is bounded by the lower and upper bounds
in~(\ref{eqn.bound1})-~(\ref{eqn.bound4})

\vspace{0.1in}

\begin{figure}[htp]
\begin{center}
\includegraphics[width=120mm]{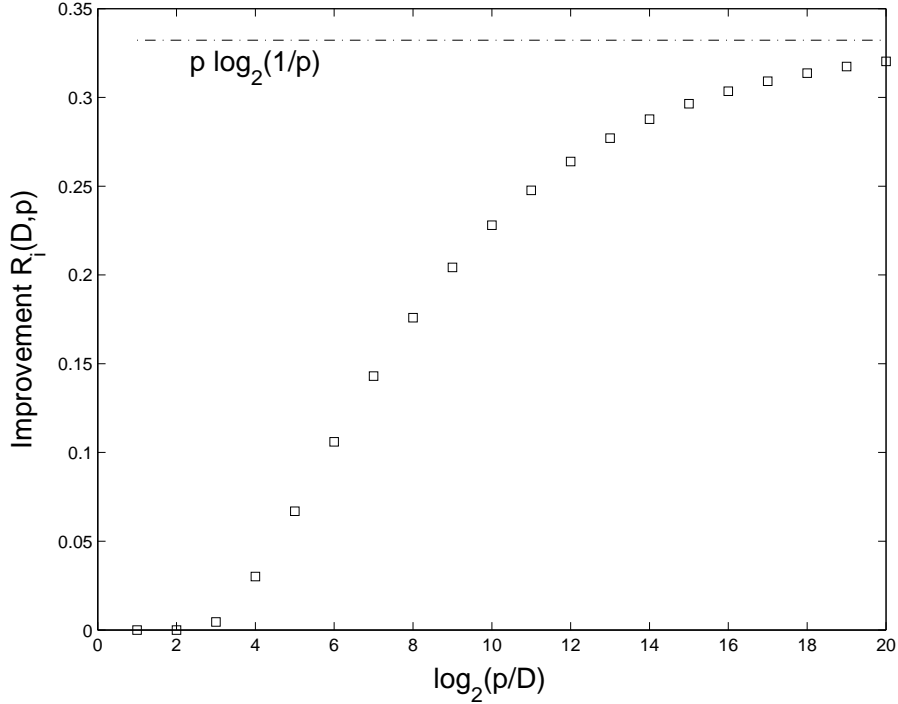}
\end{center}
\caption[ ]{ The improvement $R_i(D,p)$ for $p=0.1$ at low
distortion levels. As proved in Corollary~\ref{cor.asy},
$R_i(D,p)\rightarrow p\log_2{\frac{1}{p}}$ as $D\rightarrow 0$}
    \label{fig.Numerical1}
\end{figure}
\vspace{0.1in}

\section{Conclusions and Future Work}

In this paper we study the rate distortion function for
Bernoulli-Gaussian sequences. The main result is an improved lower
bound on the rate distortion function. The improvement over the
known best lower bound is $p\log_2\frac{1}{p}$ if $D$ is small. This
is significant since the currently known gap between the lower bound
and upper bound is $H(p)$, hence the improved lower bound is almost
tight for sparse signals where $p\ll 1$. To derive this lower bound,
we develop a new technique to lower bound part of the rate
distortion function through a randomized lossy coding channel.
 This is, to our knowledge, the first work on this topic. This new lower bound and the
 obvious upper bounds do not match. The lower bounding
 technique we use in this paper can be improved if we can relax the near-zero error
 probability constraint on the randomized channel coding. A potentially
 useful direction is to replace the channel coding part with a lossy source
 coder. This is left for future
 work.   There is another
interesting result we developed on the way to prove the main result.
We showed the equivalence of the rate distortion functions in strong
sense and expected distortion sense for continuous random variables
with finite variances.

\section*{Acknowledgments}
The author thanks Marcelo Weinberger for introducing the problem,
Erik  Ordentlich for pointing out~\cite{Fletcher_IEEE} and other
members of the information theory group at Hewlett-Packard Labs,
Palo, Alto for helpful discussions along the way.

\bibliographystyle{plain}

\bibliography{./RD_main}
\appendix

\subsection{Rate distortion function in the strong sense for continuous random variables}
\label{sec.appendix.strong}

It is shown that the rate distortions function for both the average
distortion and the strong distortion are the same for discrete
random variables Chapter 13.6~\cite{Cover}. However it is not
obvious if it is also true for continuous random variables. In this
section, we give a sketch on why it is also true for
continuous(mixed) random variables. Since we have not seen similar
results in the classic literature on rate distortion
function~\cite{Berger_lossy98}~\cite{Berger}
and~\cite{Quantization}, we feel it is necessary to give a sketch of
proof here.

\begin{figure}[htbp]
\begin{center}
     \includegraphics[width=120mm]{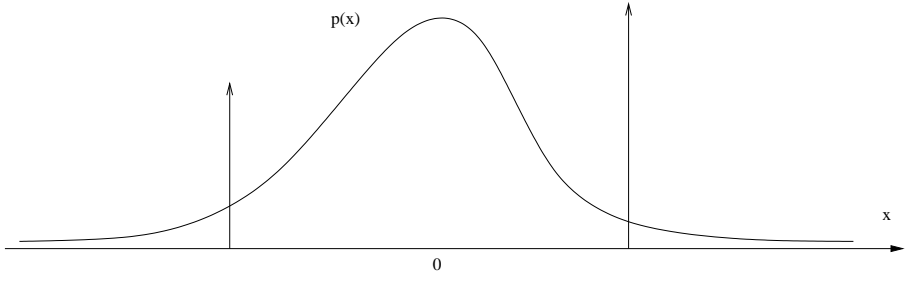}
     \end{center}
    \caption{Probability density function $p(x)$ of a continuous random variable $\rvx$}
    \label{fig:pdf}
\end{figure}

As shown in Figure~\ref{fig:pdf},  to make it more general, we let
$\rvx$ be a mixture of a continuous probability function $p(x)$ and
finite many discrete values with positive probabilities
($\Pr(\rvx=a_i)=p_i>0$ shown as impulses in the figure). We need the
mean and the variance of $\rvx$ to be finite: $E(\rvx)<\infty$ and
$E(\rvx^2)<\infty$.

First, we argue that the rate distortion function in the expected
distortion sense exists for the mixed random variables by
approximating the impulses in the pdf by a sharp step
function\footnote{For an impulse $\Pr(\rvx=a_i)=p_i>0$, we add the
continuous pdf $p(x)$ by the following step function $p_i(x)$:
$p_i(x)=m$ if $x\in [a_i-\frac{1}{2m},a_i+\frac{1}{2m}]$, $p_i(x)=0$
otherwise. } so we have a continuous pdf and the rate distortion
theorem can be applied. It remains to be shown that the continuous
rate distortion function converges to the one for $\rvx$ as
$m\rightarrow \infty$. This can be easily proved by noticing that
the approximation error is at most $\frac{1}{2m}$ for this
approximation, hence the rate distortion function of the continuous
random variable converges to the mixed one.

Now we show that the rate distortion function in the strong sense
 for continuous(mixed) random variable $\rvx$, denoted by $R_S(D,\rvx)$   is equal
to the rate distortion function in the expected sense, denoted by
$R_E(D,\rvx)$.

\begin{figure}[htbp]
\begin{center}
     \includegraphics[width=120mm]{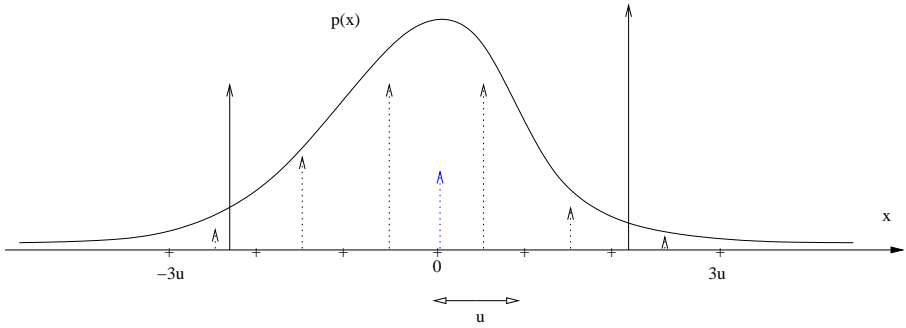}
     \end{center}
    \caption{Quantization of a probability density function $p(x)$ of a mixed random variable $\rvx$,  $7$ level quantization
    for the continuous part and exact representation of the discrete part.}
    \label{fig:pdf_Q}
\end{figure}

As shown in Figure~\ref{fig:pdf_Q}, for the continuous part of the
probability density function, we quantize the real line into
$(2K+1)$ quantization levels with the interval size $d$. The
intervals are: $[-Ku, -(K-1)u],...,[-u,0],[0,u],...,[(K-1)u, Ku]$
and the ``tail'' interval $(-\infty, -Ku]\bigcup [Ku, \infty)$. For
each interval, the representation value is the middle point of the
interval, specifically for the ``tail'' interval, the representation
value is  $0$. We use the following function $q_{K,u}$ to map a
mixed random variable to a discrete random variable:
\begin{eqnarray*}
q_{K,u}(x)= \begin{array}{ccc}
x, & p_\rvx(x)>0,\\
(k+\frac{1}{2})u,& \ \  p_\rvx(x)=0 \mbox{ and }  x\in [ku, (k+1)u),\ \ k=-K,...,K-1\\
0 ,& \ \  p_\rvx(x)=0 \mbox{ and } x\in (-\infty, -Ku]\bigcup [Ku,
\infty)
\end{array}
\end{eqnarray*}

For a random variable $\rvx$, the output of the map
$\rvy_{K,u}=q_{K,u}(\rvx)$ is a discrete random variable. Hence we
know that the rate distortion functions in the strong sense, denoted
by $R_S(D,\rvy_{K,u})$ and the expected distortion sense, denoted by
$R_E(D,\rvy_{K,u})$, are the same.

Now we have four rate distortion functions, the rate distortion
function for the mixed (continuous) random variable $\rvx$,
$R_S(D,\rvx)$ and $R_E(D,\rvx)$, and the rate distortion functions
for the quantized discrete random variables $R_S(D,\rvy_{K,u})$ and
$R_E(D,\rvy_{K,u})$. \textbf{The goal is to show that
$R_S(D,\rvx)=R_E(D,\rvx)$}. First, from the discussion in
Appendix~\ref{sec.appendix.constructing}, we know that
$R_S(D,\rvx)\geq R_E(D,\rvx)$. It remains to be shown that
$R_S(D,\rvx)\leq R_E(D,\rvx)$. We will use the discrete random
variable $\rvy_{K,u}$'s rate distortion functions as bridges to show
that. We will show that when $u\rightarrow 0$ and $Ku\rightarrow
\infty$: $R_S(D,\rvx)\leq R_S(D,\rvy_{K,u})$ and
$R_E(D,\rvy_{K,u})\leq R_E(D,\rvx)$. And knowing that for discrete
random variables $\rvy_{K,u}$,
$R_S(D,\rvy_{K,u})=R_E(D,\rvy_{K,u})$. We will have:
$$R_S(D,\rvx)\leq R_S(D,\rvy_{K,u})=R_E(D,\rvy_{K,u})\leq
R_E(D,\rvx).$$

This will conclude our proof that $R_S(D,\rvx)=R_E(D,\rvx)$. Now we
only need to show that $R_S(D,\rvx)\leq R_S(D,\rvy_{K,u})$ and
$R_E(D,\rvy_{K,u})\leq R_E(D,\rvx)$.

\subsubsection{$R_S(D,\rvx)\leq R_S(D,\rvy_{K,u})$}
We only need to show that if at a rate-distortion pair $(R,D)$,
there is a good lossy coder $\tilde f_{n_{K,u}}, \tilde g_{n_{K,u}}$
in the strong sense for $\rvy_{K,u}$, then there is a good lossy
coder $ f_n, g_n$ in the strong sense for $\rvx$.

From the definition of the good lossy coder in the strong sense, we
know that for any $\epsilon>0$,:
\begin{eqnarray}
\lim_{n\rightarrow \infty}\Pr^{\rvy_{K,u} }\left(d(\rvy_{K,u}^n,
 \tilde g_{n_{K,u}}( \tilde f_{n_{K,u}}(\rvy_{K,u}^n)))\geq  D+\delta_0
 \right)=0\nonumber
\end{eqnarray}
Notice that $\rvy_{K,u}^n= q_{K,u}(\rvx)$, so the above equation
becomes:

\begin{eqnarray}
\lim_{n\rightarrow \infty}\Pr^{\rvx }\left(d(q_{K,u}(\rvx^n),
 \tilde g_{n_{K,u}}( \tilde f_{n_{K,u}}(q_{K,u}(\rvx^n))))\geq  D+\delta_0
 \right)=0\label{eqn.appendix.partA0}
\end{eqnarray}
where the quantizer $q_{K,u}(\cdot)$ is illustrated in
Figure~\ref{fig:pdf_Q}. Now we show the following encoder decoder
pair $f_{n_{K,u}}, g_{n_{K,u}}$ is good in the strong sense for
$\rvx$ when $u$ goes to zero and $Ku$ goes to infinity. Where
$$f_{n_{K,u}}(\cdot)= \tilde f_{n_{K,u}}(q_{K,u}(\cdot)), \mbox{ and } g_{n_{K,u}}(\cdot)=\tilde g_{n_{K,u}}(\cdot).$$
Notice that the distortion $d(\cdot,\cdot)$ is the mean square of
the difference, so almost surely:
\begin{eqnarray}
d( \rvx^n ,
   g_{n_{K,u}}(   f_{n_{K,u}}( \rvx^n )))&=&d( \rvx^n ,
 \tilde g_{n_{K,u}}( \tilde
 f_{n_{K,u}}(q_{K,u}(\rvx^n))))\nonumber\\
 &\leq& d( \rvx^n ,
  q_{K,u}(\rvx^n))+d( q_{K,u}(\rvx^n) ,
 \tilde g_{n_{K,u}}( \tilde
 f_{n_{K,u}}(q_{K,u}(\rvx^n))))\nonumber\\
 &=& \frac{1}{n}\sum_{i=1}^n (\rvx_i-  q_{K,u}(\rvx_i))^2+d( q_{K,u}(\rvx^n) ,
 \tilde g_{n_{K,u}}( \tilde
 f_{n_{K,u}}(q_{K,u}(\rvx^n))))\label{eqn.appendix.partA1}
\end{eqnarray}
We analyze the first term in~(\ref{eqn.appendix.partA1}). We
decompose the sum square depending on how $\rvx_i$ is quantized,
remember for $\rvx>Ku$, the quantization is $0$ and we assume that
$Ku$ is big enough that no discrete part of $\rvx$ is larger than
$Ku$:
\begin{eqnarray}
\frac{1}{n}\sum_{i=1}^n (\rvx_i-  q_{K,u}(\rvx_i))^2&=&
\frac{1}{n}\sum_{i: |\rvx_i|\leq Ku}(\rvx_i-
q_{K,u}(\rvx_i))^2+\frac{1}{n} \sum_{i: |\rvx_i|> Ku} \rvx_i ^2
\nonumber\\
&\leq& u+ \frac{1}{n} \sum_{i: |\rvx_i|> Ku} \rvx_i ^2\nonumber
\end{eqnarray}
Pick $u< \delta_0$ and $Ku$ big enough such that $E_\rvx(1(|\rvx|>
Ku)\rvx ^2)< \delta_0-u$, this is clearly doable because $E_\rvx(
\rvx ^2)<\infty$. Now we use the weak law of large numbers,

\begin{eqnarray}
\Pr^\rvx(\frac{1}{n}\sum_{i=1}^n (\rvx_i-
q_{K,u}(\rvx_i))^2>\delta_0)&\leq&
\Pr^\rvx( \frac{1}{n} \sum_{i: |\rvx_i|> Ku} \rvx_i ^2>\delta_0-u)
\nonumber\\
&=& \Pr^\rvx( \frac{1}{n} \sum_{i=1}^n (1(|\rvx|>
Ku)\rvx ^2 >\delta_0-u)\nonumber\\
&\rightarrow&0 \ \ \ \ \  \mbox{    as $n\rightarrow \infty$
}\label{eqn.appendix.partA2}
\end{eqnarray}

Now we can bound the following probability:

\begin{eqnarray}
\Pr^\rvx(d( \rvx^n ,
   g_{n_{K,u}}(   f_{n_{K,u}}( \rvx^n )))>2\delta_0)&\leq&
\Pr^\rvx( \frac{1}{n}\sum_{i=1}^n (\rvx_i-  q_{K,u}(\rvx_i))^2+d(
q_{K,u}(\rvx^n) ,
 \tilde g_{n_{K,u}}( \tilde
 f_{n_{K,u}}(q_{K,u}(\rvx^n))))>2\delta_0)\label{eqn.appendix.partA3}
 \\
&\leq&\Pr^\rvx( \frac{1}{n}\sum_{i=1}^n (\rvx_i- q_{K,u}(\rvx_i))^2>
\delta_0)\nonumber\\
&& \ \ \ \ \ \ \ +\Pr^\rvx( d( q_{K,u}(\rvx^n) ,
 \tilde g_{n_{K,u}}( \tilde
 f_{n_{K,u}}(q_{K,u}(\rvx^n))))>\delta_0)\label{eqn.appendix.partA4}\\
&\rightarrow& 0  \ \ \ \mbox{as $n\rightarrow
\infty$}\label{eqn.appendix.partA5}
\end{eqnarray}
where (\ref{eqn.appendix.partA3})
follows~(\ref{eqn.appendix.partA1}). (\ref{eqn.appendix.partA4}) is
true because $\Pr(\rvx+\rvy>2\epsilon_0)\leq
\Pr(\rvx>\epsilon_0\mbox{ or } \rvy>\epsilon_0)\leq
\Pr(\rvx>\epsilon_0)+\Pr( \rvy>\epsilon_0)$, while
(\ref{eqn.appendix.partA5}) follows (\ref{eqn.appendix.partA0})
and~(\ref{eqn.appendix.partA2}).\\

\subsubsection{$R_E(D,\rvy_{K,u})\leq R_E(D,\rvx)$ }
We only need to show that if at a rate-distortion pair $(R,D)$,
there is a good lossy coder $f_n, g_n$ in the expected distortion
sense for $\rvx$, then there is a good lossy coder $ \tilde
f_{n_{K,u}}, \tilde g_{n_{K,u}} $ in the strong sense for
$\rvy_{K,u}$.

From the definition of the good lossy coder in the expected
distortion sense, we know that
\begin{eqnarray}
\lim_{n\rightarrow \infty} E\left(d(\rvx^n,
g_n(f_n(\rvx^n)))\right)\leq D\nonumber
\end{eqnarray}

Now we construct a good lossy coder in the expected distortion
sense, we implement the following ``inverse'' map of $q_{K,u}$,
denoted by $\rvw_{K,u}$. Where $\rvw_{K,u}$ is a random map, for any
real sequence $y^n$ generated by the random variable $\rvy_{K,u}$,
$y_i$ can only take values on $A= \{ku: k=-K,...,0,..., K \mbox{ and
}
 a\in \mathcal R \mbox{ where }p_\rvx(a)>0$, the inverse map $\rvw_{K,u}: A\rightarrow \mathcal
 R$, such that: $\rvw_{K,u}(\rvy_{K,u})\sim \rvx$ and for all $y\in
 A$: $\rvw_{K,u}(y)\in \{x\in \mathcal R: q_{K,u}(x)=y\}$.
 Pictorically  the inverse map maps the impulses in Figure~\ref{fig:pdf_Q} back to the mixed
 random variable with probability density function in
 Figure~\ref{fig:pdf}.
The good lossy coder in the expected distortion sense for
$\rvy_{K,u}$ is for all $y^n\in A^n$:
\begin{eqnarray}
&&\tilde f_{n_{K,u}}(y^n) = f_n(\rvw_{K,u}(y^n))\nonumber\\
 &&\tilde g_{n_{K,u}}=g_n\nonumber
\end{eqnarray}
Now we analyze the expected distortion of such coder.
\begin{eqnarray}
 E\left(d(\rvy_{K,u}^n, \tilde g_{n_{K,u}}(\tilde
 f_{n_{K,u}}(\rvy_{K,u}^n)))\right) &=&
 E\left(d(\rvy_{K,u}^n,   g_n(f_n(\rvw_{K,u}(\rvy_{K,u}^n))))\right)\nonumber\\
 &\leq&
 E\left(d(\rvy_{K,u}^n,   \rvw_{K,u}(\rvy_{K,u}^n))\right)+E\left(d(\rvw_{K,u}(\rvy_{K,u}^n),
 g_n(f_n(\rvw_{K,u}(\rvy_{K,u}^n))))\right)\label{eqn.appendix_construction2.1}
\end{eqnarray}

The second term in~(\ref{eqn.appendix_construction2.1}) converges to
$D$ as $n$ goes to infinity because $\rvw_{K,u}(\rvy_{K,u}^n))\sim
\rvx^n$ and $f_n,g_n$ is good for $\rvx^n$ in the expected
distortion sense.  As for the first term in
~(\ref{eqn.appendix_construction2.1}), we show it converges to zero
for small $u$ and big $Ku$ as $n$ goes to infinity.
\begin{eqnarray}
  E_{\rvy_{K,u}}\left(d(\rvy_{K,u}^n,   \rvw_{K,u}(\rvy_{K,u}^n))\right)&=&E_{\rvy_{K,u}}(\frac{1}{n}\sum_{i=1}^n
 (\rvy_{K,u}(i)-\rvw_{K,u}(\rvy_{K,u}(i)))^2)\nonumber\\
 &=&E_{\rvy_{K,u}}(
 (\rvy_{K,u} -\rvw_{K,u}(\rvy_{K,u} ))^2)\nonumber\\
 &=& E_{\rvy_{K,u}}( 1(\rvw_{K,u}(\rvy_{K,u} )\leq
Ku)
 (\rvy_{K,u} -\rvw_{K,u}(\rvy_{K,u} ))^2)\nonumber\\
&& \ \ \ \ \ \ \ + E_{\rvy_{K,u}}(
 1(\rvw_{K,u}(\rvy_{K,u} )>
Ku)(\rvy_{K,u} -\rvw_{K,u}(\rvy_{K,u} ))^2)\nonumber\\
 &\leq & \frac{u^2}{4} + E_{\rvy_{K,u}}(
 1(\rvw_{K,u}(\rvy_{K,u} )>
Ku)(\rvy_{K,u} -\rvw_{K,u}(\rvy_{K,u} ))^2)\label{eqn.appendix_construction2.2}\\
 &=& \frac{u^2}{4} + E_{\rvx}(
 1(\rvx>
Ku) \rvx^2)\label{eqn.appendix_construction2.3}\\
&\rightarrow& 0 \mbox{ as  } u\rightarrow 0 \mbox{ and }
Ku\rightarrow \infty\label{eqn.appendix_construction2.4}
 \end{eqnarray}
(\ref{eqn.appendix_construction2.2}) is true because if
$|\rvw_{K,u}(\rvy_{K,u} )|\leq Ku$, then the quantization error is
no bigger than $\frac{u}{2}$. (\ref{eqn.appendix_construction2.3})
follows that $\rvw_{K,u}(\rvy_{K,u} )\sim\rvx$.
(\ref{eqn.appendix_construction2.4}) is true because the variance of
$\rvx$ is finite. (\ref{eqn.appendix_construction2.4}) and
(\ref{eqn.appendix_construction2.1}) gives us the desired result
that the expected distortion of $\tilde g_{n_{K,u}},\tilde
 f_{n_{K,u}}$ converges to $D$ if $u$ goes to zero, $Ku$ goes to
 infinity.

\subsection{Constructing a good lossy source coding in the expected distortion sense from a good one in the strong sense}
\label{sec.appendix.constructing}

The construction here is a general proof. It works for both
continuous, discrete and mixed random variables. By constructing a
good lossy source coder in the expected distortion sense from a good
lossy coder in the strong sense at the same rate-distortion point
$(R,D)$, we can easily see that the rate distortion function in the
strong sense is not smaller than the rate distortion function in the
expected distortion sense. This fact is used in the proof in
Appendix~\ref{sec.appendix.strong}.

Assume both the first and second order moment of $\rvx$ are finite,
i.e. $E(\rvx)=\mu_\rvx<\infty$ and $E(\rvx^2)=\sigma_\rvx<\infty$.
If $f_n, g_n$ is good in the strong sense for $R(D)$, then we denote
by $\Upsilon_n\subseteq\mathcal R^n$, the subset the distortion
constraint is not satisfied, i.e. $\Upsilon_n=\{x^n\in \mathcal R^n:
d(\svx^n, g_n(f_n(\svx^n))\geq D+\delta\}$. Denote by
$e_n=\Pr\limits^{\rvx}(\Upsilon_n)$, then $e_n\rightarrow 0$. A good
lossy coder might have $g_n(f_n(\svx^n))$ arbitrarily faraway from
$\svx^n$ for $\svx^n\in \Upsilon_n$ as pointed out in
Section~\ref{sec.intro_R(D)} and cause the expected distortion
arbitrarily large. We build a new lossy coding system $\tilde f_n,
\tilde g_n$, such that $\tilde g_n(\tilde
f_n(\svx^n))=g_n(f_n(\svx^n))$ for $\svx^n\notin \Upsilon_n$ and
$\tilde g_n(\tilde f_n(\svx^n))=0$ for $\svx^n\in \Upsilon_n$.
Obviously is good in the strong sense, we only need to show that
$\tilde f_n, \tilde g_n$ is also good in the expected distortion
sense. The expected distortion of $\tilde f_n, \tilde g_n$ is:
\begin{eqnarray}
E(d(\rvx^n, \tilde g_n(\tilde f_n(\rvx^n))))&=&\Pr(\rvx^n\in
\Upsilon_n^C ) E(d(\rvx^n, \tilde g_n(\tilde f_n(\rvx^n)))|\rvx^n\in
\Upsilon^C_n)+\Pr(\rvx^n\in \Upsilon_n ) E(d(\rvx^n, \tilde
g_n(\tilde f_n(\rvx^n)))|\rvx^n\in
\Upsilon_n)\nonumber\\
&\leq &(1-e_n)(D+\delta)+\Pr(\rvx^n\in \Upsilon_n )E(d(\rvx^n,
\tilde g_n(\tilde
f_n(\rvx^n)))|\rvx^n\in \Upsilon_n)\nonumber\\
&= &(1-e_n)(D+\delta)+\Pr(\rvx^n\in \Upsilon_n
)E(\frac{1}{n}\sum_{i=1}^n \rvx_i^2|\rvx^n\in
\Upsilon_n)\label{eqn.sgtrongtoexpect1}
\end{eqnarray}
Now we upper bound the second term, first according to the weak law
of large numbers and the variance and the mean of $\rvx$ are finite,
we know that for any $\epsilon>0$, there exists $n_\epsilon<\infty$,
s.t for all $n> n_\epsilon$:
\begin{eqnarray}
\Pr^{\rvx}(|\frac{1}{n}\sum_{i=1}^n \rvx_i^2
-\sigma_\rvx|>\epsilon)<\epsilon.
\end{eqnarray}
This implies that for any subset $\Gamma\in \mathcal R^n$ with
measure $\Pr\limits^\rvx(\Gamma)\geq 1-\epsilon$, then there is a
subset $ \Gamma_1\subseteq \Gamma$, such that
$\Pr\limits^\rvx(\Gamma_1)\geq 1-2\epsilon$ and  for all $\svx^n\in
\Gamma_1$: $|\frac{1}{n}\sum_{i=1}^n \svx_i^2 -\sigma_\rvx|\leq
\epsilon$.

From the definition of $e_n$, we know that for large enough $n$,
$e_n=\Pr\limits^{\rvx}(\Upsilon_n)< \epsilon$ or equivalently
$\Pr\limits^{\rvx}(\Upsilon^C_n)\geq 1-\epsilon$. From the above
discussion, there exists subset $\Gamma_1\in\Upsilon^C_n$, such that
$\Pr\limits^{\rvx}(\Gamma_1)\geq 1-2\epsilon$ and  for all
$\svx^n\in \Gamma_1$: $|\frac{1}{n}\sum_{i=1}^n \svx_i^2
-\sigma_\rvx|\leq \epsilon$.  So the expectation of the mean
variance of $\rvx^n$ can be decomposed:

\begin{eqnarray}
\sigma_\rvx&=& E( \frac{1}{n}\sum_{i=1}^n \rvx_i^2
)\nonumber\\
&=&\Pr(\rvx^n\in \Upsilon^C_n ) E(\frac{1}{n}\sum_{i=1}^n
\rvx_i^2|\rvx^n\in \Upsilon^C_n)+\Pr(\rvx^n\in \Upsilon_n )
E(\frac{1}{n}\sum_{i=1}^n \rvx_i^2|\rvx^n\in
\Upsilon_n)\nonumber\\
&\geq&\Pr(\rvx^n\in \Gamma_1 ) E(\frac{1}{n}\sum_{i=1}^n
\rvx_i^2|\rvx^n\in \Gamma_1)+\Pr(\rvx^n\in \Upsilon_n )
E(\frac{1}{n}\sum_{i=1}^n \rvx_i^2|\rvx^n\in
\Upsilon_n)\nonumber\\
&\geq&(1-\epsilon)(\sigma_\rvx-\epsilon)+\Pr(\rvx^n\in \Upsilon_n )
E(\frac{1}{n}\sum_{i=1}^n \rvx_i^2|\rvx^n\in \Upsilon_n)\nonumber
\end{eqnarray}
Hence:
\begin{eqnarray}
\Pr(\rvx^n\in \Upsilon_n ) E(\frac{1}{n}\sum_{i=1}^n
\rvx_i^2|\rvx^n\in \Upsilon_n)\leq
\epsilon(1+\sigma_\rvx)\label{eqn.sgtrongtoexpect2}
\end{eqnarray}
Substituting (\ref{eqn.sgtrongtoexpect1})
into~(\ref{eqn.sgtrongtoexpect2}), we have:
\begin{eqnarray}
E(d(\rvx^n, \tilde g_n(\tilde
f_n(\rvx^n))))\leq(1-e_n)(D+\delta)+\epsilon(1+\sigma_\rvx)\leq
D+\delta+\epsilon(1+\sigma_\rvx)\nonumber
\end{eqnarray}

Note that  the above is true for all $\epsilon$ and $\delta$, so we
can let both be arbitrarily small and the expected distortion of
$\tilde f_n, \tilde g_n$ is arbitrarily close to $D$.  Hence we just
constructed a good lossy coding system in the expected distortion
sense from a good lossy coding system in the strong sense. \hfill
$\square$

\subsection{Proof of the simple bounds:
proof of Propositions~\ref{prop.equavalence}, \ref{prop.upper1},
~\ref{prop.upper2} and~\ref{prop.lower_trivial} }
\label{sec.proof_upper}

\textbf{Proof of Proposition~\ref{prop.equavalence}:}

To show $R(D, \Xi(p,\sigma^2))\geq R(\frac{D}{\sigma^2}, \Xi(p,
 1))$, we only need to construct a sequence of good, in the strong
sense of rate distortion in (\ref{eqn.defn_rd_strong}),
encoder/decoder pairs $(f'_n, g'_n), \ n=1,2,...$, for $\Xi(p, 1)$
from that for $\Xi(p, \sigma^2)$, $(f_n, g_n), \ n=1,2,...$. Let
$f'_n$ and $g'_n$ be as follows, for all $x^n\in \mathcal X^n$ and
$a^{nR}\in \{0,1\}^{nR}$:
\begin{eqnarray}
f'_n(x^n)=f_n(\sigma{x^n}), \ \ \ g'_n(a^{nR})=\frac{1}{\sigma}
g_n(a^{nR})\nonumber
\end{eqnarray}
So for $\rvx\sim\Xi(p, 0,1)$
\begin{eqnarray}
\Pr\left(d(\rvx^n, g'_n(f'_n(\rvx^n)))\geq \frac{D+\delta}{\sigma^2}
\right) &=& \Pr\left(d(\rvx^n,
\frac{1}{\sigma}g_n(f_n(\sigma\rvx^n)))\geq
 \frac{D+\delta}{\sigma^2}\right)\nonumber\\
 & =& \Pr \left(d({\sigma}\rvx^n, g_n(f_n(\sigma\rvx^n)))\geq
  {D+\delta}  \right)\label{eqn.transform1}
\end{eqnarray}
where~(\ref{eqn.transform1}) is because the distortion measure $d(x,
y)=(x-y)^2$ in this paper.

 Obviously for  $\rvx\sim\Xi(p, 1) $, $
{\sigma}\rvx\sim\Xi(p,\sigma^2)$, and if $f_n$ and $g_n$ are good in
the strong sense, defined in~(\ref{eqn.defn_rd_strong}), for $\Xi(p,
\sigma^2)$, then for all $\delta>0$:
\begin{eqnarray}
\lim_{n\rightarrow \infty} \Pr\left(d(\sigma \rvx^n, g_n(f_n( \sigma
\rvx^n))\geq D+\delta \right)=0\label{eqn.transform2}.
\end{eqnarray}

Combining~(\ref{eqn.transform1}) and~(\ref{eqn.transform2}), we
have:
\begin{eqnarray}
\lim_{n\rightarrow \infty} \Pr\left(d(\rvx^n,
g'_n(f'_n(\rvx^n)))\geq \frac{D+\delta}{\sigma^2}
\right)=0\nonumber.
\end{eqnarray}

Notice that $\delta$ is an arbitrary positive number and $\sigma$ is
constant, we just show that $R(D, \Xi(p,\sigma^2))\geq
R(\frac{D}{\sigma^2}, \Xi(p,
 1))$. Similarly we can show that $R(D, \Xi(p, \sigma^2))\leq
R(\frac{D}{\sigma^2}, \Xi(p, 1))$. This complete the proof that
$R(D, \Xi(p,\sigma^2))= R(\frac{D}{\sigma^2}, \Xi(p, 1))$.
\hfill$\square$

\vspace{0.1in}

\textbf{Proof of Proposition ~\ref{prop.upper1}:} for a
Bernoulli-Gaussian random sequence $\rvx^n$, by
Definition~\ref{def.bernoulli-gaussian}, we know that
$\rvx_i=\rvb_i\times \rvs_i$, $\rvb_i\sim Bernoulli-p$ and
$\rvs_i\sim N(0,1)$ are i.i.d random variables. The encoder $f_n$
works as follows. It is consisted of two parts. First the encoder
encode $\rvb^n$ \textit{losslessly} using  a fixed length code-book.
Then the encoder encode \textit{lossily} the subsequence of $\rvs^n$
where $\rvb_i\neq 0$ by applying standard Gaussian lossy source
coding.

We now describe the coding scheme $f_n, g_n$, in details. If $b^n$
is $\epsilon_1$-strong typical,  and write $1(b^n)$ as the number of
$1$'s in sequence $b^n$. i.e.:
\begin{eqnarray}
b^n \in B^n_{\epsilon_1}\triangleq\{b^n\in \{0,1\}^n:
|\frac{1(b^n)}{n}-p|\leq \epsilon_1\}.\nonumber
\end{eqnarray}
then $f_n$ one-to-one maps $b^n$ to a binary sequence of length
$n(H(p)+\tau(\epsilon_1))$ excluding the all zero signal, otherwise
$b^n \notin B^n_{\epsilon_1}$,
  $f_n$ sends the all zero signal, where
$\tau(\epsilon_1)\rightarrow 0$ if $\epsilon_1\rightarrow 0$, this
is guaranteed by the standard lossless source coding theorem.
Obviously for all $\epsilon_1>0$:
\begin{eqnarray}
\lim_{n\rightarrow \infty} \Pr^\rvb(\rvb^n\notin
B^n_{\epsilon_1})=0\label{eqn.appendix.probabilityB}
\end{eqnarray}
Now for each $\svx^n=b^n\times s^n$, if $b^n \in B^n_{\epsilon_1}$,
we know that $n(p-\epsilon_1)\leq 1(b^n)\leq n(p+\epsilon_1)$.
Denote by a new sequence $\tilde s_1,... \tilde s_{1(b^n)}$ the non
zero entries of $\svx^n$. Then the encoder $f_n$ passes $\tilde
s^{1(b^n)}$ to a good Gaussian lossy encoder-decoder pair $\tilde
f_{1(b^n)}, \tilde g_{1(b^n)}$ with rate $R(\frac{D}{p}, N(0,1))$
for a sequence of length $1(b^n)$. If output of  $\tilde
f_{1(b^n)}$, when $1(b^n)< n(p+\epsilon_1)$, is shorter than
$n(p+\epsilon_1) R(\frac{D}{p}, N(0,1))$, $f_n$ just pad zeros at
the end. The total block length for $x^n$ is
\begin{eqnarray}
n(H(p)+\tau(\epsilon_1)) +n(p+\epsilon_1) R(\frac{D}{p},
N(0,1)).\label{eqn.appendix.total_length}
\end{eqnarray}

If the output form the encoder is not a all zero sequence, the
decoder $g_n$ first looks at the first $n(H(p)+\tau(\epsilon_1))$
bits and recover $b^n$ exactly and hence $1(b^n)$. Then  $g_n$
discards the padded zeros at the end and pass the rest to the
Gaussian lossy decoder $\tilde g_{1(b^n)}$ with rate $R(\frac{D}{p},
N(0,1))$ for a sequence of length $1(b^n)$. Then $g_n$ put the
outputs of $\tilde g_{1(b^n)}$ to the non-zero locations of $b^n$
one by one. By using the coding system described above, we have for
$b^n\in B^n_{\epsilon_1}$,
\begin{eqnarray}
n d(\svx^n, g_n(f_n(\svx^n))) =1(b^n) d(\tilde \svs^{1(b^n)}, \tilde
g_{1(b^n)}(\tilde f_{1(b^n)}(\tilde
\svs^{1(b^n)})))\label{eqn.appendix.tilde1}
\end{eqnarray}
and because $\rvs^n$ and $\rvb^n$ are independent and the coding
system $\tilde f_{1(b^n)}, \tilde g_{1(b^n)}$ is good, for all fixed
$b^n\in B^n_{\epsilon_1}$, for all $\delta_0>0$:
\begin{eqnarray}\lim_{n\rightarrow \infty}\Pr^{\tilde \rvs}\left(d(\tilde \rvs^{1(b^n)},
 \tilde g_{1(b^n)}( \tilde f_{1(b^n)}(\tilde \rvs^{1(b^n)})))\geq \frac{D}{p}+\delta_0
 \right)=0.\label{eqn.appendix.tilde_S}
\end{eqnarray}

Now we evaluate the performance of $f_n, g_n$, for all $\delta_1>0$:
\begin{eqnarray}
\lim_{n\rightarrow \infty}\Pr^\rvx\left(d(\rvx^n,
g_n(f_n(\rvx^n)))\geq D+\delta_1\right)&\leq & \lim_{n\rightarrow
\infty}\{\Pr^\rvb\left(\rvb^n\notin B^n_{\epsilon_1}\right) +
\Pr^\rvx\left(d(\rvx^n, g_n(f_n(\rvx^n)))\geq D+\delta_1|\rvb^n\in
B^n_{\epsilon_1}\right)\}\label{eqn.appendix.prop2.0}\\
&=&  \lim_{n\rightarrow \infty}  \Pr^\rvx\left(d(\rvx^n,
g_n(f_n(\rvx^n)))\geq D+\delta_1|\rvb^n\in
B^n_{\epsilon_1}\right)\label{eqn.appendix.prop2.1}\\
&=&  \lim_{n\rightarrow \infty}  \Pr^\rvx\left(d(\tilde
\rvs^{1(b^n)},
 \tilde g_{1(b^n)}( \tilde f_{1(b^n)}(\tilde \rvs^{1(b^n)})))\geq \frac{n(D+\delta_1)}{ 1(\rvb^n)}|\rvb^n\in B^n_{\epsilon_1}\right)\label{eqn.appendix.prop2.2}\\
&\leq &  \lim_{n\rightarrow \infty}  \Pr^\rvx\left(d(\tilde
\rvs^{1(b^n)},
 \tilde g_{1(b^n)}( \tilde f_{1(b^n)}(\tilde \rvs^{1(b^n)})))\geq \frac{n(D+\delta_1)}{ n(p+
 \epsilon_1)}|\rvb^n\in B^n_{\epsilon_1}\right)\label{eqn.appendix.prop2.3}\\
&= &  \lim_{n\rightarrow \infty}  \Pr^\rvx\left(d(\tilde
\rvs^{1(b^n)},
 \tilde g_{1(b^n)}( \tilde f_{1(b^n)}(\tilde \rvs^{1(b^n)})))\geq \frac{ D }{ p}+\frac{\delta_1 p-D \epsilon_1}{p(p+\epsilon_1)}|\rvb^n\in B^n_{\epsilon_1}\right)\nonumber\\
&=&0\label{eqn.appendix.prop2.4}
 \end{eqnarray}
(\ref{eqn.appendix.prop2.0}) is because for events $A$ and $B$,
$\Pr(A)=\Pr(A, B)+\Pr(A,B^c)\leq \Pr( B)+\Pr(A|B^c)$.
(\ref{eqn.appendix.prop2.1}) is true because
of~(\ref{eqn.appendix.probabilityB}). (\ref{eqn.appendix.tilde1})
implies (\ref{eqn.appendix.prop2.2}). (\ref{eqn.appendix.prop2.3})
is true because $1(b^n) \leq n(p+\epsilon_1)$ if $b^n\in
B^n_{\epsilon_1}$. Finally, for any $\delta_1$, by letting
$\epsilon_1$ small enough, hence $\frac{\delta_1 p-D
\epsilon_1}{p(p+\epsilon_1)}>0$ and by~(\ref{eqn.appendix.tilde_S})
and the fact that $\tilde \rvs^{1(\rvb^n)}$ is induced by $\rvx^n$,
we have~(\ref{eqn.appendix.prop2.4}).

(\ref{eqn.appendix.total_length}) together
with~(\ref{eqn.appendix.prop2.4}) implies that:
\begin{eqnarray}
R(D,p)\leq H(p)+p R(\frac{D}{p}, N(0,1))+\tau(\epsilon_1)
+\epsilon_1R(\frac{D}{p}, N(0,1)).\nonumber
\end{eqnarray}
Notice that we can pick $\epsilon_1$ arbitrarily small and hence
$\tau(\epsilon_1)$ arbitrarily small, we have $R(D,p)\leq
H(p)+pR(\frac{D}{p}, N(0,1))= H(p)+pR({D}, N(0,p))$. \hfill$\square$
\vspace{0.1in}

\textbf{Proof of Proposition ~\ref{prop.upper2}:} this is a direct
corollary of the upper bound in
Corollary~\ref{cor.rate_cover_bounds}. Notice that the variance of a
Bernoulli-Gaussian random variable $ \Xi(p, 1)$ is $p$, so according
to Corollary~\ref{cor.rate_cover_bounds}, if $ \Xi(p, 1)$ is a
continuous random variable, we would have:
\begin{eqnarray}
R(D, p)\leq \max\{\frac{1}{2}\log
 \frac{p}{D},0\}= R(D, N(0,p))
\end{eqnarray}

The technicality here is that $\Xi(p, 1)$ is not a continuous random
variable, but the fix is quite easy. Let random variable $\rvy_m$ be
the $p$-mixture of a Gaussian $N(0,1)$ and a uniformly distributed
random variable on $[-\frac{1}{m},\frac{1}{m}]$. i.e. with
probability $1-p$, $\rvy_m\sim N(0,1)$ and with probability $p$,
$\rvy_m\sim U[-\frac{1}{m},\frac{1}{m}]$. The pdf of $\rvy_m$ is
\begin{eqnarray}
p_{\rvy_m}(y)=\{ \begin{array}{ccc}
\frac{1-p}{2}e^{\frac{y^2}{2}}, & |y|> \frac{1}{m},\\
\frac{1-p}{2}e^{\frac{y^2}{2}}+ \frac{pm}{2},& \ \  |y|\leq
\frac{1}{m}.
\end{array}
\end{eqnarray}

Obviously, $\rvy_m$ is a continuous random variable with variance
$p+\frac{1}{3m^2}$. Now according to
Corollary~\ref{cor.rate_cover_bounds}, we know that the rate
distortion function for $\rvy_m$, $R_{\rvy_m}(D)$ is upper bounded
by \begin{eqnarray}\max\{\frac{1}{2}\log
 \frac{p}{D},0\}= R(D,
 N(0,p+\frac{1}{3m^2})).\label{eqn.appendix.upperboundym}
 \end{eqnarray}

 Now we upper bound $R(D,p)$ by constructing a good randomized lossy coding
 system for $\rvx\sim \Xi(p, 1)$ in the average sense,
  $f^\rvu_n, g^\rvu_n$,  from a good lossy coding system for
 $\rvy_m$, $f^\rvy_n, g^\rvy_n$. Given $\rvx^n\sim\Xi(p, 1)$, $f^\rvu_n$
 applies  the following operation on it, for $i=1,2,..., n $, let
\begin{eqnarray}
\rvz_i=\{ \begin{array}{ccc}
\rvx_i, & \rvx_i\neq 0,\\
\rvx_i+\rvu_i,& \ \  \rvx_i=0
\end{array}
\end{eqnarray}
where $\rvu_i$'s are independent and $\rvu_i\sim U[\frac{-1}{m},
\frac{1}{m}]$ is a uniform distributed random variable. It is clear
that $\rvz_i$ has the same distribution as $\rvy_m$. Now $f^\rvu_n$
passes $\rvz^n$ to encoder $f^\rvy_n$. The decoder
$g^\rvu_n=g^\rvy_n$. Now we analyze the performance of the coding
system $f^\rvu_n, g^\rvu_n$.

First, because $f^\rvy_n, g^\rvy_n$ is a good in the strong sense
for $\rvy_m$, we have, for any $\delta_1>0$:
\begin{eqnarray}
\lim_{n\rightarrow \infty}\Pr^\rvz\left(d(\rvz^n,
g^\rvy_n(f^\rvy_n(\rvz^n))\geq D+\delta_1 \right)=0.\nonumber
\end{eqnarray}
From the construction of $f^\rvu_n,g^\rvu_n$, we know that
$g^\rvy_n(f^\rvy_n(\rvz^n))=g^\rvu_n(f^\rvu_n(\rvx^n))$ a.s., where
$\rvz^n$ is induced from $\rvx^n$ and $\rvu^n$,  denote
$g^\rvu_n(f^\rvu_n(\rvx^n))_i$ or equivalently
$g^\rvy_n(f^\rvy_n(\rvz^n))_i$  by $\rvw_i$, so :
\begin{eqnarray}
\lim_{n\rightarrow \infty}\Pr^{\rvx,\rvu}\left(d(\rvz^n,
g^\rvu_n(f^\rvu_n(\rvx^n))\geq D+\delta_1
)\right)=\lim_{n\rightarrow
\infty}\Pr^{\rvx,\rvu}\left(d(\rvz^n,\rvw^n)\geq D+\delta_1
)\right)=0.\label{eqn.appendix.temp0}
\end{eqnarray}

 Secondly, from the
construction of $\rvz^n$, we know that for all $i$, $
|\rvx_i-\rvz_i|\leq \frac{1}{m}$ a.s..  So we have a.s.:
\begin{eqnarray}
d(\rvx^n, g^\rvu_n(f^\rvu_n(\rvx^n))) &=& \frac{1}{n}\sum_{i=1}^n (\rvx_i-\rvw_i)^2 \nonumber\\
&=& \frac{1}{n}\sum_{i=1}^n (\rvx_i-\rvz_i+\rvz_i- \rvw_i)^2\nonumber\\
&\leq& \frac{1}{n}\sum_{i=1}^n (|\rvx_i-\rvz_i|+|\rvz_i- \rvw_i|)^2\nonumber\\
&\leq& \frac{1}{n}\sum_{i=1}^n (\frac{1}{m}+|\rvz_i- \rvw_i|)^2\nonumber\\
&\leq& \frac{1}{m}+\frac{1}{n}\sum_{i=1}^n (\rvz_i- \rvw_i)^2+
\frac{2}{nm}\sum_{i=1}^n |\rvz_i- \rvw_i|\label{eqn.appendix.temp1}
\end{eqnarray}
By the Cauchy-Schwartz inequality, a.s.:
\begin{eqnarray}
(\sum_{i=1}^n |\rvz_i- \rvw_i|)^2\leq (\sum_{i=1}^n |\rvz_i-
\rvw_i|^2) (\sum_{i=1}^n 1)\nonumber
\end{eqnarray}
hence a.s.
\begin{eqnarray}
\frac{1}{n}\sum_{i=1}^n |\rvz_i- \rvw_i|\leq
\sqrt{\frac{1}{n}(\sum_{i=1}^n |\rvz_i- \rvw_i|^2)
}\label{eqn.appendix.temp2}.
\end{eqnarray}
Now for a realization of $\rvx^n$ and $\rvu^n$: $\svx^n$ and
$\svu^n$,  the induced realization of $\rvz^n$ and
$g^\rvu_n(f^\rvu_n(\rvx^n))$ are $\svz^n$ and  $\svw^n$
respectively.  If $d(\svz^n, \svw^n)< D+\delta_1$, then
combining~(\ref{eqn.appendix.temp1}) and~(\ref{eqn.appendix.temp2}),
we have:
\begin{eqnarray}
d(\svx^n, g^\svu_n(f^\svu_n(\svx^n)))
&\leq& \frac{1}{m}+(D+\delta_1)+ \frac{2\sqrt{D+\delta_1}}{
m} \nonumber\\
&\leq&  D+\delta_1 + \frac{1+2\sqrt{D+\delta_1}}{ m}\nonumber
\end{eqnarray}
This means that
\begin{eqnarray}
\Pr^{\rvx,\rvu}\left(d(\rvz^n, \rvw^n)\geq D+\delta_1 \right) \geq
\Pr^{\rvx,\rvu}\left(d(\rvx^n, \rvw^n)\geq D+\delta_1 +
\frac{1+2\sqrt{D+\delta_1}}{ m}\right)\label{eqn.appendix_temp3}
\end{eqnarray}
The above coding system is a randomized coding system where the
performance is measured under the distribution of the ``dithering''
random variable $\rvu$. Now if we take the above average
``dithering'', i.e.: for each $x^n\in R^n$,

$$ \mbox{if} \ \ \ \ \ \ \ \ \ \ \Pr^{ \rvu}\left(d(\svx^n, \rvw^n)\geq D+\delta_1 +
\frac{1+2\sqrt{D+\delta_1}}{ m}\right)<1,$$

 there exists $u^n(x^n)\in [\frac{-1}{m}, \frac{1}{m}]$ and of course $u_i(x^n)=0$
if $x_i\neq 0$, such that the distortion between $x^n$ and the
output of the the lossy coding system $f^\rvy_n, g^\rvy_n$ with
input $ x^n +u^n(x^n)$ is no bigger than  $ D+\delta_1 +
\frac{1+2\sqrt{D+\delta_1}}{ m}$:
\begin{eqnarray}
d(\svx^n, g^\rvy_n(f^\rvy_n(\svx^n+\svu^n(\svx^n))))\leq D+\delta_1
+ \frac{1+2\sqrt{D+\delta_1}}{ m}\nonumber
\end{eqnarray}
$$\mbox{otherwise, simply let }  \ \ \ \ \ \ \ \ \ \ \  u^n(x^n)=0.$$

Finally, let $f_n$ and $g_n$ be such that, for all $x^n$, $
f_n(x^n)=f^\rvy_n(\svx^n+\svu^n(\svx^n))$ and $g_n=g^\rvy_n$. The
construction of $g_n, f_n$ implies that
\begin{eqnarray}
\Pr^\rvx(d(\rvx^n, g_n(f_n(\rvx^n))\geq D+\delta_1 +
\frac{1+2\sqrt{D+\delta_1}}{ m})\leq \Pr^{\rvx,\rvu}\left(d(\rvx^n,
\rvw^n)\geq D+\delta_1 + \frac{1+2\sqrt{D+\delta_1}}{ m}\right)
\label{eqn.appendix_temp4}
\end{eqnarray}

Now combining~(\ref{eqn.appendix.temp0}),~(\ref{eqn.appendix_temp3})
and~(\ref{eqn.appendix_temp4}), we have:
\begin{eqnarray}
\lim_{n\rightarrow \infty}\Pr^\rvx(d(\rvx^n, g_n(f_n(\rvx^n))\geq
D+\delta_1 + \frac{1+2\sqrt{D+\delta_1}}{ m})=0\nonumber
\end{eqnarray}
Note that the rate of the coding system is $R_{\rvy_m}(D)$ which is
upper bounded by $R(D, N(0,p+\frac{1}{3m^2}))$
in~(\ref{eqn.appendix.upperboundym}). So
 \begin{eqnarray}
R(D+ \delta_1 + \frac{1+2\sqrt{D+\delta_1}}{ m},p)\leq R(D,
 N(0,p+\frac{1}{3m^2}))\label{eqn.appendix.final}
 \end{eqnarray}
 while~(\ref{eqn.appendix.final}) is true for all $\delta_1>0$ and
 $m\in \mathcal N$. Note that the Gaussian  rate distortion function $R(D,
 N(0,\sigma^2))$ is continuous in $\sigma^2$ and the
 Bernoulli-Gaussian rate distortion function $R(D,p)$ is
 monotonically decreasing and bounded in $D$, hence continuous with
 measure $1$. By letting $m\rightarrow \infty$ and $\delta_1\rightarrow 0$, we have:
 $R(D,p)\leq R(D, N(0,p))$.  \hfill$\square$

\vspace{0.1in}

\textbf{Proof of Proposition ~\ref{prop.lower_trivial}:} For a good
lossy coding system $f_n, g_n$ for Bernoulli-Gaussian sequence
$\rvx^n= \rvb^n\times \rvs^n \sim \Xi(p,1)$ defined in
Definition~\ref{def.bernoulli-gaussian} and distortion constraint
$D$, the rate is $R(D,p)$, i.e.
$$f_n: \mathcal \mathcal R^n \rightarrow \{0,1\}^{n R(D,p)}\ \
\ \  g_n:  \{0,1\}^{n R(D,p)} \rightarrow \mathcal R^n, \ \ \
\Pr^\rvx\left(d(\rvx^n, g_n(f_n(\rvx^n)))\geq
D+\delta_1\right)=e_n$$
\begin{eqnarray}
\mbox{and for all $\delta_1>0$:     } \lim_{n\rightarrow
\infty}e_n=0.\label{eqn.appendix.goodbernoulligaussian}
\end{eqnarray}

We use the same notations as those in the proof of
Proposition~\ref{prop.upper1}. We construct a good length $m_n\in
[n(p-\epsilon_1), n(p+\epsilon_1)]$ lossy source coding system
$\tilde f_{m_n}, \tilde g_{m_n}$ for $\tilde \rvs^{{m_n}}\sim
N(0,p)$ under the same distortion constraint $D$, where $m_n$ will
be determined later. First we decompose $e_n$,
by~(\ref{eqn.appendix.probabilityB}), we know that there exists
$n_{\epsilon_1}<\infty$, such that for all $n> n_{\epsilon_1}$,
$\Pr\limits^\rvx(\rvb^n\in B^n_{\epsilon_1})\geq \frac{1}{2}$, so
for all $n> n_{\epsilon_1}$:

\begin{eqnarray}
e_n&=&\Pr^\rvx\left(d(\rvx^n, g_n(f_n(\rvx^n)))\geq
D+\delta_1\right)\nonumber\\
&\geq &  \Pr^\rvx\left(\rvb^n \in B^n_{\epsilon_1} \mbox{ , }
d(\rvx^n, g_n(f_n(\rvx^n)))\geq
D+\delta_1\right)\label{eqn.appendix_proposition4.temp1}\\
&= & \sum_{b^n\in B^n_{\epsilon_1}} \Pr^\rvx\left(\rvb^n=b^n \mbox{
, } d(\rvx^n, g_n(f_n(\rvx^n)))\geq
D+\delta_1\right)\label{eqn.appendix_proposition4.temp2}\\
&= & \Pr\limits^\rvx(\rvb^n \in B^n_{\epsilon_1})\sum_{  b^n \in
B^n_{\epsilon_1}} \frac{\Pr\limits^\rvx(\rvb^n=b^n
)}{\Pr\limits^\rvx(\rvb^n \in B^n_{\epsilon_1})}\Pr^\rvx\left(
d(\rvx^n, g_n(f_n(\rvx^n)))\geq D+\delta_1| \rvb^n=b^n
\right)\label{eqn.appendix_proposition4.temp3}\\
&\geq &\frac{1}{2} \sum_{  b^n \in B^n_{\epsilon_1}} \phi(b^n)
\Pr^\rvx\left( d(\rvx^n, g_n(f_n(\rvx^n)))\geq D+\delta_1|
\rvb^n=b^n \right)\label{eqn.appendix_proposition4.temp4}
\end{eqnarray}
(\ref{eqn.appendix_proposition4.temp1}),~(\ref{eqn.appendix_proposition4.temp2})
and~(\ref{eqn.appendix_proposition4.temp3}) are obvious,
in~(\ref{eqn.appendix_proposition4.temp4}), we denote $\phi(b^n)$ by
$ \frac{\Pr\limits^\rvx(\rvb^n=b^n )}{\Pr\limits^\rvx(\rvb^n \in
B^n_{\epsilon_1})}$. Notice that $\phi()$ is  a probability measure
on $B^n_{\epsilon_n}$. Hence  there exists $ \bar b^n\in \mathcal
B^n$, write $1(\bar b^n)=m_n\in [n(p-\epsilon_1), n(p+\epsilon_1)]$,
such that:
\begin{eqnarray}
\Pr^\rvx\left( d(\rvx^n, g_n(f_n(\rvx^n)))\geq D+\delta_1|
\rvb^n=\bar b^n \right)\leq 2
e_n\label{eqn.appendix_proposition4.temp5}.
\end{eqnarray}
We bound the distortion of $\svx^n$ as follows, let $l_1<l_2<...<
l_{m_n}$,  $L=\{l_1,...l_{m_n}\}$,  be the positions of the non-zero
elements of $\bar b^n$,
\begin{eqnarray}
n d(\svx^n, g_n(f_n(\svx^n)))\geq \sum_{i=1}^{m_n}
(x_{l_i}-g_n(f_n(\svx^n))_{l_i})^2\label{eqn.appendix_proposition4.temp6}.
\end{eqnarray}
Substituting~(\ref{eqn.appendix_proposition4.temp6})
into~(\ref{eqn.appendix_proposition4.temp5}), we have:
\begin{eqnarray}
\Pr^\rvx\left( \sum_{i=1}^{m_n}
(\rvx_{l_i}-g_n(f_n(\rvx^n))_{l_i})^2\geq n( D+\delta_1)|
\rvb^n=\bar b^n \right)\leq 2
e_n\label{eqn.appendix_proposition4.temp7}.
\end{eqnarray}
  Now we are ready to
construct a good lossy source coding system $\tilde f_{m_n}, \tilde
 g_{m_n}$ for $\rvs^{m_n}\sim N(0,p)$. The encoder $\tilde f_{m_n}$
works as follows, for any sequence $\tilde \svs^{m_n} \in \mathcal
R^{m_n}$, $\tilde f_{m_n}(\tilde \svs^{m_n})= f_n(T(\svs^{m_n}))$,
for a binary sequence $a^{R(D,p)n}\in\{0,1\}^{R(D,p)n}$: $\tilde
g_{m_n}(a^{R(D,p)n})= T^{-1}g_n(a^{R(D,p)n}))$, where $T$ is a
one-to-one map from $\mathcal R^{m_n}$ to $\mathcal R^{n}$:
\begin{eqnarray}
&& T(\tilde \svs^{m_n})=\svs^n, \mbox{ where } \svs_{l_i}=\tilde
\svs_i, \ \ i=1,2,..., m_n \mbox{ and } \svs_i=0, \ i\notin
L\nonumber\\
&& T^{-1}(\svs^n)=\tilde \svs^{m_n}, \mbox{ where } \tilde
\svs_i=\svs_{l_i}, \ \ i=1,2,..., m_n\nonumber
\end{eqnarray}

$\rvx^n=\rvb^n \times \rvs^n$, so if $\rvb^n=\bar b^n$ then
$\rvx_i=0$ for all $i\notin L$, and by the memorylessness  of
$\rvx^n$. We have:
\begin{eqnarray}
\Pr^{\tilde \rvs}\left(m_n d(\rvs^{m_n}, \tilde g_{m_n}(\tilde
f_{m_n}(\rvs^{m_n})) )\geq n(D+\delta_1) \right)&=&\Pr^{\tilde
\rvs}\left( \sum_{i=1}^{m_n} (\tilde \rvs_{i}-\tilde g_{m_n}(\tilde
f_{m_n}(\tilde \rvs^n))_{i})^2\geq n(D+\delta_1) \right)\nonumber\\
&=&\Pr^\rvx\left( \sum_{i=1}^{m_n}
(\rvx_{l_i}-g_n(f_n(\svx^n))_{l_i})^2\geq  n(D+\delta_1)|
\rvb^n=\bar
b^n \right)\nonumber\\
&\leq& 2 e_n\label{eqn.appendix_proposition4.temp8}.
\end{eqnarray}
where the inequality is by~(\ref{eqn.appendix_proposition4.temp7}).
Notice that $m_n=1(\tilde b^n)\in
[n(p-\epsilon_1),n(p+\epsilon_1)]$, so $\frac{n}{m_n}\in
[\frac{1}{p+\epsilon_1}, \frac{1}{p-\epsilon_1}]$.
So~(\ref{eqn.appendix_proposition4.temp8})
and~(\ref{eqn.appendix.goodbernoulligaussian}) tells us:

\begin{eqnarray}
0&=& \lim_{n\rightarrow \infty}\Pr^{\tilde \rvs}\left(m_n
d(\rvs^{m_n}, \tilde g_{m_n}(\tilde f_{m_n}(\rvs^{m_n})) )\geq
n(D+\delta_1) \right) \nonumber\\
&=& \lim_{n\rightarrow \infty}\Pr^{\tilde \rvs}\left( d(\rvs^{m_n},
\tilde g_{m_n}(\tilde f_{m_n}(\rvs^{m_n})) )\geq
\frac{n}{m_n}(D+\delta_1) \right) \nonumber\\
&\geq & \lim_{n\rightarrow \infty}\Pr^{\tilde \rvs}\left(
d(\rvs^{m_n}, \tilde g_{m_n}(\tilde f_{m_n}(\rvs^{m_n})) )\geq
\frac{1}{p-\epsilon_1}(D+\delta_1) \right)
\label{eqn.appendix_proposition4.temp9}.
\end{eqnarray}
The encoder decoder pair $\tilde f_{m_n}, \tilde g_{m_n}$ use
$nR(D,p)$ bits, so the rate of this coding system is
$\frac{nR(D,p)}{m_n}\leq \frac{R(D,p)}{p-\epsilon_1}$.
(\ref{eqn.appendix_proposition4.temp9}) is true for all $\delta_1$
and $\epsilon_1$, by letting $\epsilon_1\rightarrow 0$, we just
construct a rate $\frac{R(D,p)}{p}$, distortion $\frac{D}{p}$ coding
system for i.i.d Gaussian random variables $\tilde \rvs^{m_n}\sim
N(0,1)$. From Corollary~\ref{cor.rate_gaussian} we know that
$\frac{R(D,p)}{p} \geq R(\frac{D}{p}, N(0,1))$, i.e.
$$ {R(D,p)}  \geq p R(\frac{D}{p}, N(0,1))= p R({D}, N(0,p ))$$

\hfill$\square$

\subsection{ Strong Typical Gaussian Sequences
}\label{sec.strong_typical_gaussian}

In this appendix we define and investigate the properties of the so
called strong typical Gaussian sequences. For a sequence $\svs^n\in
\mathcal R^n$, for a real number $T\in \mathcal R$, the empirical
$l$-th moment of entries in $\svs^n$ within interval  $[T, \infty]$
is denoted by
$$n^l_{\svs^n}(T)=\frac{\sum_{i=1}^n 1(\svs_i>T)\svs_i^l}{n} .$$

\begin{definition}{ $\epsilon$-typical Gaussian sequences:}\label{def.etypical} A
sequence $\svs^n$ is said to be $\epsilon$ typical for $N(0,1)$, if
the followings are true: for any  real number  $T\geq -\infty$,
\begin{eqnarray}
\max_{l=0,1,2}\left\{\sup_{T}\left|n^l_{\svs^n}(T)-\int_T^{\infty}\svs^l\frac{1}{\sqrt{2\pi}}e^{\frac{-s^2}{2}}ds\right|<\epsilon\right\}
\end{eqnarray}
The $\epsilon$-typical set of $N(0,1)$ is denoted by
$S_{\epsilon}(n)$, similar to the strong typical set for random
sequences with finite alphabet, we have the following concentration
lemma. Note that the convergence is
uniform convergence, in the sense that we ask the  sequence to be typical for all real numbers $T$ simultaneously.\\

An almost equivalent ``double-sided'' definition of
$\epsilon$-typical Gaussian sequence is as follows.  First, for any
$-\infty\leq S\leq T\leq \infty$, we denote by
$$n^{l*}_{\svs^n}(S, T)=\frac{\sum_{i=1}^n 1(\svs_i\in [S,T])\svs_i^l}{n }.$$

Similar to that in Definition~\ref{def.etypical}, we define the
typical set $S^*_{\epsilon}(n)$ as the set of all sequence $s^n$,
s.t.

\begin{eqnarray}
\max_{l=0,1,2}\left\{\sup_{S\leq
T}\left|n^*_{\svs^n}(S,T)-\int_S^{T}\svs^l\frac{1}{\sqrt{2\pi}}e^{\frac{-s^2}{2}}ds\right|<\epsilon
\right\}\label{eqn.typicality}
\end{eqnarray}
We now illustrate the equivalence of the two typical sets
$S_{\epsilon}(n)$ and $S^*_{\epsilon}(n)$. First, obviously
$S^*_{\epsilon}(n)\subseteq S_{\epsilon}(n)$. Secondly,
\begin{eqnarray}
\sup_{S\leq
T}\left|n^{l*}_{\svs^n}(S,T)-\int_S^{T}\svs^l\frac{1}{\sqrt{2\pi}}e^{\frac{-s^2}{2}}ds\right|&=
& \sup_{S\leq
T}\left|n^l_{\svs^n}(S)-n^l_{\svs^n}(T)-\int_S^{\infty}\svs^l\frac{1}{\sqrt{2\pi}}e^{\frac{-s^2}{2}}ds+
\int_T^{\infty}\svs^l\frac{1}{\sqrt{2\pi}}e^{\frac{-s^2}{2}}ds\right|\nonumber\\
&\leq & \sup_{S\leq
T}\left|n^l_{\svs^n}(S)-\int_S^{\infty}\svs^l\frac{1}{\sqrt{2\pi}}e^{\frac{-s^2}{2}}ds\right|+\left|
n^l_{\svs^n}(T)-\int_T^{\infty}\svs^l\frac{1}{\sqrt{2\pi}}e^{\frac{-s^2}{2}}ds\right|\nonumber\\
&\leq & 2 \sup_{
T}\left|n^l_{\svs^n}(T)-\int_T^{\infty}\svs^l\frac{1}{\sqrt{2\pi}}e^{\frac{-s^2}{2}}ds\right|\nonumber
\end{eqnarray}
This means $S_{\epsilon}(n) \subseteq S^*_{2\epsilon}(n)$, so the
concentration of the ``double-sided'' and the ``one-sided'' typical
sets are equivalent. We use the latter definition of
$\epsilon$-typical set in the main body of the paper. However, for
the sake of simplicity of notations, we prove the concentration of
the $\epsilon$-typical set of the
``one-sided'' definition.\\

\begin{lemma}{Concentration of Gaussian sequences:}\label{lemma.concentrationGaussian}
for i.i.d $N(0,1)$ random sequence $\rvs^n$, for all $\epsilon>0$
\begin{eqnarray}
\lim_{n\rightarrow \infty} \Pr(\rvs^n\in S_\epsilon(n))=1
\end{eqnarray}

\end{lemma}
\vspace{0.1in}
 \proof we give a sketch of the proof here. The idea
is to first quantize the real line for the Gaussian $N(0,1)$ random
variable then apply the concentration result for i.i.d discrete
finite random sequences. The quantization goes as follows, we study
the following intervals: $\{(-\infty, -K\omega], [-K\omega,
-(K-1)\omega], ... ,[(K-1)\omega, K\omega], [K\omega, \infty]\}$,
i.e. the end points of the intervals are defined as follows: for an
integer $j$ within range $[-K-1, K+1]$ we denote $\omega(j)=
j\omega$ if $j=-K,...,K$ and $\omega(-K-1)=-\infty$ and
$\omega(K+1)=\infty$. We can obviously let $\omega$ be small enough
and $K$ be big enough such that the following two integrals are true
for all $j=-K-1,...K$
\begin{eqnarray}
  \left|\int_{\omega(j)}^{
\omega(j+1)}s^l\frac{1}{\sqrt{2\pi}}e^{\frac{-s^2}{2}}ds\right|<\frac{\epsilon}{2}\mbox{
for } l =0,1,2.\label{eqn.definition_EP_T}
\end{eqnarray}
We let $S^{\omega, K}_\epsilon(n)$ be the set  that the typicality
condition in~(\ref{eqn.typicality}) is true for $T=\omega(j)$ for
all $j\in\{-K-1,...,K+1\}$ simultaneously, i.e.

\begin{eqnarray}
S^{\omega, K}_\epsilon(n)=\{\svs^n:
\max_{l=0,1,2}\left\{\sup_{j=-K-1,...,K+1}\left|n^l_{\svs^n}(\omega(j))-\int_{\omega(j)}^{\infty}\svs^l\frac{1}{\sqrt{2\pi}}e^{\frac{-s^2}{2}}ds\right|<\epsilon
\right\}\}\label{eqn.typicality0.5}
\end{eqnarray}
We show that
\begin{eqnarray}
\lim_{n\rightarrow \infty} \Pr(\rvs^n\in S^{\omega,
K}_\epsilon(n))=1\label{eqn.typical_quantized}
\end{eqnarray}
This is true because from the weak law of large numbers we know that
for $l=0,1,2$:
\begin{eqnarray}
\lim_{n\rightarrow \infty} \Pr
\left(\left|n^l_{\rvs^n}(T)-\int_T^{\infty}\svs^l\frac{1}{\sqrt{2\pi}}e^{\frac{-s^2}{2}}ds\right|<\epsilon\right)=1
\label{eqn.typicality1}
\end{eqnarray}
for all $T\in \mathcal R\bigcup \{-\infty, \infty\}$, in particular
for all $T=\omega(j)$, $j=-K-1,..., K+1$. This is a finite set, so
\begin{eqnarray}
&&\lim_{n\rightarrow \infty} \Pr(\rvs^n\in S^{\omega,
K}_\epsilon(n))\nonumber\\
&&=\lim_{n\rightarrow \infty} \Pr
 (
 \max_{l=0,1,2}\left\{\sup_{j=-K-1,...,K+1}\left|n^l_{\svs^n}(\omega(j))-\int_{\omega(j)}^{\infty}\frac{1}{\sqrt{2\pi}}e^{\frac{-s^2}{2}}ds\right|\right\}<\epsilon)\nonumber\\
 && =1\label{eqn.typicality2}
\end{eqnarray}
(\ref{eqn.typical_quantized}) is proved. In particular:
\begin{eqnarray}
\lim_{n\rightarrow \infty} \Pr(\rvs^n\in S^{\omega,
K}_{\frac{\epsilon}{2}}(n))=1\label{eqn.typical_quantized1}
\end{eqnarray}

 Now we are ready to
use~(\ref{eqn.typical_quantized1}) to prove the lemma.

For any $\svs^n$ and  a real number $T\in [\omega(j), \omega(j+1)]$,
$j\in\{-K-1,..., K+1\}$, then obviously $$n^l_{\svs^n}(T)\in
[n^l_{\svs^n}(\omega(j+1)), n^l_{\svs^n}(\omega(j))], \mbox{ so for
} l=0,1,2:
$$
\begin{eqnarray}
n^l_{\svs^n}(T)-\int_T^{\infty}\svs^l\frac{1}{\sqrt{2\pi}}e^{\frac{-s^2}{2}}ds
&\leq& n^l_{\svs^n}(\omega(j))-
\int_T^{\infty}\svs^l\frac{1}{\sqrt{2\pi}}e^{\frac{-s^2}{2}}ds\nonumber\\
&=&
n^l_{\svs^n}(\omega(j))-\int_{\omega(j)}^{\infty}\svs^l\frac{1}{\sqrt{2\pi}}e^{\frac{-s^2}{2}}ds+
\int_{\omega(j)}^{T}\svs^l\frac{1}{\sqrt{2\pi}}e^{\frac{-s^2}{2}}ds\nonumber\\
&\leq &
n^l_{\svs^n}(\omega(j))-\int_{\omega(j)}^{\infty}\svs^l\frac{1}{\sqrt{2\pi}}e^{\frac{-s^2}{2}}ds+
\int_{\omega(j)}^{\omega(j+1)}\svs^l\frac{1}{\sqrt{2\pi}}e^{\frac{-s^2}{2}}ds\nonumber\\
&\leq &
\left|n^l_{\svs^n}(\omega(j))-\int_{\omega(j)}^{\infty}\svs^l\frac{1}{\sqrt{2\pi}}e^{\frac{-s^2}{2}}ds\right|+\frac{\epsilon}{2}\label{eqn.boundingbyT0},
\end{eqnarray}
where~(\ref{eqn.boundingbyT0}) follows~(\ref{eqn.definition_EP_T}),
similarly we have for $l=0,1,2:$
\begin{eqnarray}
n^l_{\svs^n}(T)-\int_T^{\infty}\svs^l\frac{1}{\sqrt{2\pi}}e^{\frac{-s^2}{2}}ds&\geq
&
-\left|n^l_{\svs^n}(\omega(j+1))-\int_{\omega(j+1)}^{\infty}\svs^l\frac{1}{\sqrt{2\pi}}e^{\frac{-s^2}{2}}ds\right|-\frac{\epsilon}{2}\label{eqn.boundingbyT1}.
\end{eqnarray}
(\ref{eqn.boundingbyT0}) and~(\ref{eqn.boundingbyT1}) tells us that
for $l=0,1,2:$
\begin{eqnarray}
\left|n^l_{\svs^n}(T)-\int_T^{\infty}\svs^l\frac{1}{\sqrt{2\pi}}e^{\frac{-s^2}{2}}ds\right|&\leq
&
\sup_{j=-K-1,...,K+1}\left|n^l_{\svs^n}(\omega(j))-\int_{\omega(j)}^{\infty}\svs^l\frac{1}{\sqrt{2\pi}}e^{\frac{-s^2}{2}}ds\right|+\frac{\epsilon}{2}\label{eqn.boundingbyT2}.
\end{eqnarray}

Notice the definitions of $S_{\epsilon}(n)$ and $S^{\omega,
K}_{\frac{\epsilon}{2}}(n)$, ~(\ref{eqn.boundingbyT2})
  implies that $S_{\epsilon}(n) \supseteq
S^{\omega, K}_{\frac{\epsilon}{2}}(n)$, hence:

\begin{eqnarray}
\lim_{n\rightarrow \infty} \Pr(\rvs^n\in S_\epsilon(n))&\geq&
\lim_{n\rightarrow \infty} \Pr(\rvs^n\in S^{\omega,
K}_{\frac{\epsilon}{2}}(n) )=1\nonumber
\end{eqnarray}
The lemma is proved. \hfill $\square$\\
\end{definition}

\subsection{ Properties of $(u+v) D(\frac{u}{u+v}\| p )$
}\label{sec.property_of_EE}

In this section we show some properties of $(u+v) D(\frac{u}{u+v}\|
p )$, summarized in the following lemma. \\
\begin{lemma}\label{lemma.property_of_EE} If $u, v\geq 0$, $\frac{u}{u+v}>p$,
 then $(u+v) D(\frac{u}{u+v}\| p
)$ is monotonically increasing with $u$ and monotonically decreasing
with $v$.\\
\end{lemma}

\proof First, both $\frac{u}{u+v}$ and $D(\frac{u}{u+v}\| p )$ are
positive and  monotonically increasing with $u$ if $ \frac{u}{u+v}>
p $. Hence $(u+v) D(\frac{u}{u+v}\| p )$ is monotonically increasing
with $u$.\\

Secondly, using basic calculus, we have:

\begin{eqnarray}
\frac{d (u+v) D(\frac{u}{u+v}\| p )}{d v} &=& \frac{d\left(u \log
(\frac{u}{(u+v)p})+ v \log (\frac{v}{(u+v)(1-p)})
\right)}{dv}\nonumber\\
&=&-\frac{u}{u+v}- \frac{v}{u+v}+1 +\log\left
(\frac{v}{(u+v)(1-p)}\right)\nonumber\\
&=& \log\left
(\frac{1- \frac{u}{u+v}}{1-p}\right)\nonumber\\
&<&0
\end{eqnarray}
The last inequality is true because $\frac{u}{u+v}>p$ hence
$1-\frac{u}{u+v}<1-p$. \hfill $\square$

\end{document}